\documentclass[acmsmall,screen]{acmart}

\usepackage{graphicx}
\usepackage{textcomp}
\usepackage{array}
\usepackage{epstopdf}
\usepackage{subfigure}
\usepackage{caption}
\usepackage{multirow}
\usepackage{setspace}
\usepackage{algorithm}
\usepackage{algorithmic}
\usepackage{listings}
\usepackage{xcolor}
\usepackage[misc]{ifsym}
\usepackage{listings}
\usepackage{makecell}
\usepackage{color}
\usepackage{tcolorbox}
\definecolor{dkgreen}{rgb}{0,0.6,0}
\definecolor{gray}{rgb}{0.5,0.5,0.5}
\definecolor{mauve}{rgb}{0.58,0,0.82}
\usepackage{caption}

\usepackage{url}

\usepackage{xcolor}

\newcommand{\ie}{\textit{i.e.,} }
\newcommand{\eg}{\textit{e.g.,} }

\newcommand\opt[1]{}
\newcommand\find[1]{}

\newcommand{\ls}[1]
   {\dimen0=\fontdimen6\the\font 
    \lineskip=#1\dimen0
    \advance\lineskip.5\fontdimen5\the\font
    \advance\lineskip-\dimen0
    \lineskiplimit=.9\lineskip
    \baselineskip=\lineskip
    \advance\baselineskip\dimen0
    \normallineskip\lineskip
    \normallineskiplimit\lineskiplimit
    \normalbaselineskip\baselineskip
    \ignorespaces
   }

\newenvironment{smalldescription}{
   \setlength{\topsep}{0pt}
   \setlength{\partopsep}{0pt}
   \setlength{\parskip}{0pt}
   \begin{description}
   \setlength{\leftmargin}{.2in}
   \setlength{\parsep}{0pt}
   \setlength{\parskip}{0pt}
   \setlength{\itemsep}{0pt}}{\end{description}}

\acmNumber{Preprint}

\setcopyright{none}
\renewcommand\footnotetextcopyrightpermission[1]{} 

\begin{document}

\title{Multi-Programming Language Sandbox for LLMs}

\author{\footnotesize Shihan Dou$^{1}$, Jiazheng Zhang$^{1}$, Jianxiang Zang$^{1}$, Yunbo Tao$^{1}$, Weikang Zhou$^{1}$, Haoxiang Jia$^{2}$, Shichun Liu$^{1}$, Yuming Yang$^{1}$, Zhiheng Xi$^{1}$, Shenxi Wu$^{1}$, Shaoqing Zhang$^{5}$, Muling Wu$^{1}$, Changze Lv$^{1}$, Limao Xiong$^{6}$, Wenyu Zhan$^{6}$, Lin Zhang$^{6}$, Rongxiang Weng$^{6}$, Jingang Wang$^{6}$, Xunliang Cai$^{6}$, Yueming Wu$^{3}$, Ming Wen$^{4}$, Rui Zheng$^{1}$, Tao Ji$^{1}$, Yixin Cao$^{1}$, Tao Gui$^{1}$, Xipeng Qiu$^{1}$, Qi Zhang$^{1}$, Xuanjing Huang$^{1}$}
\affiliation{%
  \institution{\\$^{1}$Fudan University, $^{2}$Peking University, $^{3}$Nanyang Technological University, $^{4}$Huazhong University of Science and Technology, $^{5}$Harbin Institute of Technology, $^{6}$Meituan Inc}
  \country{.}
}
\email{shdou21@m.fudan.edu.cn}
\authornote{Tao Gui and Xuanjing Huang are the corresponding authors.
Shihan Dou, Jiazheng Zhang, and Jianxiang Zang contributed equally. 
Jiazheng Zhang, Shaoqing Zhang, and Muling Wu are interns at Meituan LLM Team.
MLPSandbox has been used for large-scale training and various downstream code-related tasks at Meituan Inc.
Contact information: shdou21@m.fudan.edu.cn, haoxiangjia@pku.edu.cn, wengrongxiang@meituan.com, and tgui@fudan.edu.cn \\
}

\makeatletter
\let\@authorsaddresses\@empty
\makeatother

\pagestyle{fancy}
\fancyhf{} 
\fancyhead[C]{Multi-Programming Language Sandbox for LLMs} 
\fancyfoot[R]{\thepage} 
\fancyfoot[L]{FudanNLP} 

\renewcommand{\headrulewidth}{1pt} 
\renewcommand{\footrulewidth}{1pt} 

\definecolor{headrulecolor}{rgb}{0.0, 0.0, 0.5}
\definecolor{footrulecolor}{rgb}{0.0, 0.0, 0.5} 

\makeatletter
\renewcommand{\headrule}{%
    \color{headrulecolor}\hrule\@height\headrulewidth\@width\headwidth}
\renewcommand{\footrule}{%
    \color{footrulecolor}\hrule\@height\footrulewidth\@width\headwidth}
\makeatother




\begin{abstract}

We introduce MPLSandbox, an out-of-the-box multi-programming language sandbox designed to provide unified and comprehensive feedback from compiler and analysis tools for Large Language Models (LLMs). 
It can automatically identify the programming language of the code, compiling and executing it within an isolated sub-sandbox to ensure safety and stability.
In addition, MPLSandbox also integrates both traditional and LLM-based code analysis tools, providing a comprehensive analysis of generated code.
MPLSandbox can be effortlessly integrated into the training and deployment of LLMs to improve the quality and correctness of their generated code.
It also helps researchers streamline their workflows for various LLM-based code-related tasks, reducing the development cost. 
To validate the effectiveness of MPLSandbox, we integrate it into training and deployment approaches, and also employ it to optimize workflows for a wide range of real-world code-related tasks.
Our goal is to enhance researcher productivity on LLM-based code-related tasks by simplifying and automating workflows through delegation to MPLSandbox.


\end{abstract}

\maketitle

\section{Introduction}

Recently, researchers have become increasingly interested in the development of large language models (LLMs) for code-related tasks \cite{le2023codechain, shin2023prompt, pan2024lost}.
To improve the performance of LLMs in these tasks, some studies utilize sandboxes to compile and execute code, providing compiler feedback information for LLMs \cite{le2022coderl, liu2023rltf}.
However, existing sandbox tools tend to serve mono-programming languages and are also not conveniently integrated into the training and deployment processes of LLMs \cite{engelberth2012pybox, Terrarium, promptfoo, LLMSandbox}.
The lack of well-developed multi-language sandbox environments significantly limits the application of LLMs in tasks involving multiple programming languages.

Additionally, comprehensive code analysis, such as code smell, fuzz testing, and execution efficiency, assists LLMs in better understanding the code, thereby improving the quality of the generated code and enhancing their effectiveness for downstream tasks \cite{wang2022execution, lu2024grace, du2024generalization}.
However, the diversity in analytic tools and their distinct applications increases the complexity of their utilization and development for researchers. 
Moreover, the same type of analysis tool often varies across different programming languages \cite{manes2019art, gentleman2007statistical}.
In a multi-programming language context, this variation in analysis tools across different languages significantly increases the cost for researchers to develop and employ these tools to enhance the performance of models in code-related tasks.

To address these challenges, we propose MPLSandbox, an out-of-the-box sandbox designed to provide unified compiler feedback across multiple programming languages.
Additionally, it integrates traditional code analysis tools, delivering comprehensive code information to LLMs from numerous perspectives.
MPLSandbox simplifies code analysis for researchers, and can be seamlessly integrated into LLM training and application processes to enhance the performance of LLMs in a range of code-related tasks.
MPLSandbox consists of three core modules: the ``Multi-Programming Language Sandbox Environment'' which provides unified compiler feedback by compiling and executing the code; the ``Code Analysis Module'' which includes multiple traditional analysis tools to offer a comprehensive analysis report from numerous perspectives; and the ``Information Integration Module'' which integrates compilation feedback and various analysis results to accomplish a range of complex code-related tasks.
For the first module (\ie the multi-programming language sandbox environment), the code and unit test samples are sent to the sub-sandbox of the corresponding programming language for isolated execution to obtain compiler feedback. 
The sandbox ensures the program executes safely without jeopardizing the external environment or interrupting the training process \cite{Terrarium, promptfoo}. 
The second module (\ie the analysis module) provides a comprehensive code analysis from multiple perspectives, such as static analysis (\eg potential bug detection \cite{habib2018many} and code smell analysis \cite{pereira2022code, liu2019deep}) and dynamic analysis (\eg fuzz testing \cite{godefroid2008automated} and efficiency analysis \cite{berger2023triangulating, li2023djxperf}).
Additionally, this module can also assess other input information besides the code, such as evaluating the coverage of unit tests for the code, aiding researchers in improving the quality of these unit tests.
Finally, the third module (\ie the information integration module) integrates these results for LLMs to improve the quality of generated code and enhance their performance on a range of code-related tasks.

Specifically, the features of our proposed MPLSandbox include:
\begin{itemize}
    \item \textbf{Security and Stability.} 
    MPLSandbox constructs sub-sandboxes for each programming language, ensuring that programs are compiled and executed in isolation from the training environment. 
    This setup also prevents LLM-generated code that contains malicious vulnerabilities or bugs harms the external environment. 
    Additionally, various vulnerability and bug detection tools are integrated into the static analysis module to further ensure safety.
    \item \textbf{Multi-programming language support.} 
    We are the first to propose a multi-programming language sandbox. 
    MPLSandbox can automatically identify the programming language of the code, post it to the corresponding sandbox environment, and effortlessly and thoroughly analyse the code using numerous analysis tools. 
    It significantly reduces the development cost for researchers in developing and deploying LLMs for a range of code-related tasks.
    \item \textbf{Usability and Extensibility.} 
    MPLSandbox integrates multiple analysis tools for each programming language, and users can also effortlessly design tool templates to integrate their tools into MPLSandbox. 
    Furthermore, users can easily construct prompt templates to combine compiler feedback and analysis results to enhance LLM's performance in code-related tasks.
    \item \textbf{Distributed Architecture.} 
    MPLSandbox is designed for distributed deployment. 
    In large-scale training scenarios, training nodes can access the optional MPLSandbox nodes. 
    Compared to deployments where both training nodes and sandbox nodes are co-located on a single machine, MPLSandbox offers greater efficiency.
    
\end{itemize}

To validate the effectiveness of MPLSandbox, we conducted extensive experiments involving training, production, and deployment scenarios by using reinforcement learning \cite{le2022coderl}, Best-of-N (BoN) \cite{yuan2023rrhf}, and self-correction \cite{li2023reflection}. 
The experimental results demonstrate that MPLSandbox can be easily integrated into various application scenarios to enhance the correctness and quality of generated code.
In addition, we explored the application of MPLSandbox in several real-world code-related tasks, including unit test generation, bug fixing, vulnerability localization, and code translation.
By facilitating the easy integration of analysis tools and the flexible combination of different modules, MPLSandbox significantly enhances the performance of LLMs in code-related tasks, while maintaining low development costs for researchers. 
MPLSandbox is the first to offer a multi-programming language sandbox that simplifies the complexity of employing LLMs in code tasks and provides opportunities for their improvement. 
Our tool can help drive further research in this area.

\begin{figure}[htbp]
\centerline{\includegraphics[width=1\textwidth]{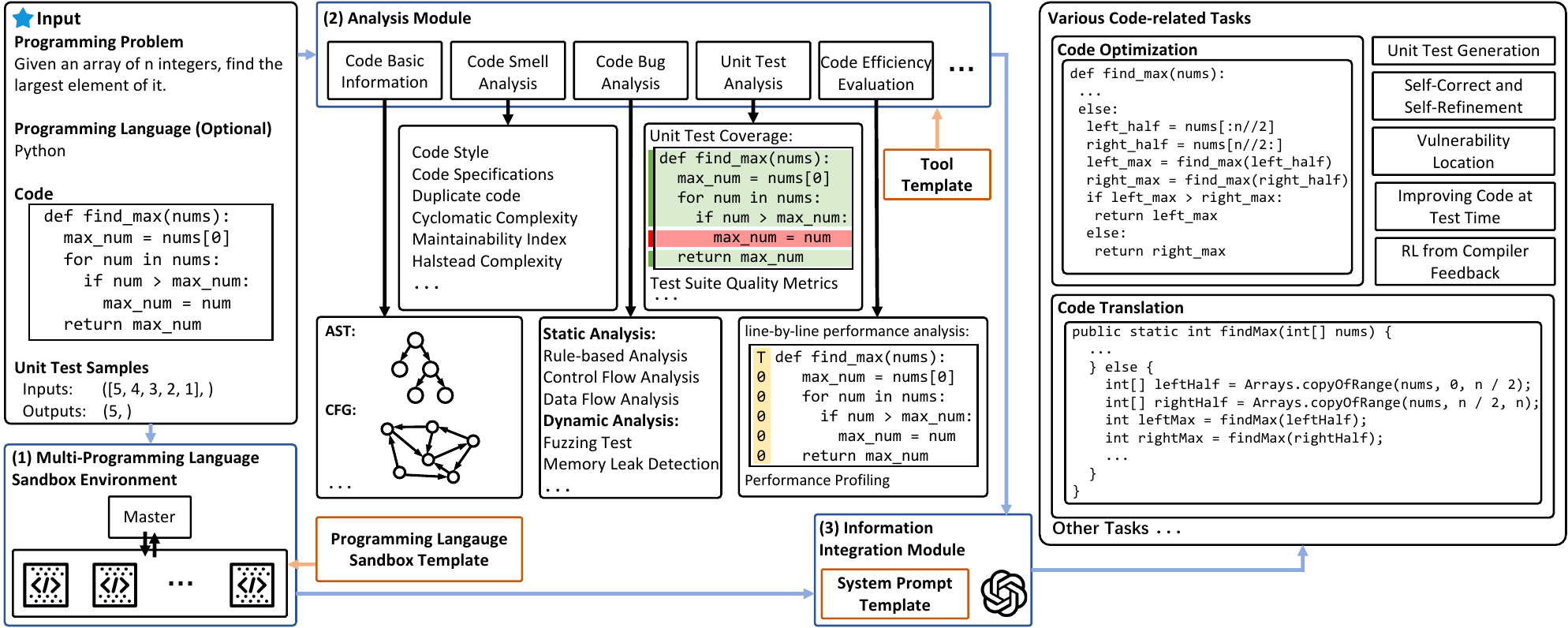}}
\caption{
The architecture of MPLSandbox.
It comprises three core modules: (1) Multi-Programming Language Sandbox Environment, (2) Code Analysis Module, and (3) Information Integration Module.
The Multi-Programming Language Sandbox Environment can provide unified compiler feedback by compiling and executing the code. 
The Code Analysis Module contains multiple traditional analysis tools to offer a comprehensive analysis report from numerous perspectives. 
The Information Integration Module integrates compilation feedback and various analysis
results to accomplish a range of complex code-related tasks.
}
\label{fig:arch}
\end{figure}

\section{MPL-Sandbox}

In this section, we introduce the architecture and pipeline of our proposed MPLSandbox.

\subsection{Architecture}

MPLSandbox is an out-of-the-box multi-programming language sandbox designed to provide unified compiler feedback and comprehensive code analysis for LLMs. 
It enables researchers to thoroughly analyze LLM-generated code in any programming language using sandbox and various code analysis tools, significantly reducing development costs. 
Additionally, MPLSandbox can streamline LLMs' training and deployment workflows for various code-related tasks.
The architecture of MPLSandbox is shown in Figure~\ref{fig:arch}.

It receives four inputs: a description of the programming problem, the code to be analyzed, the programming language used (optional), and unit test samples.
If the user or LLMs do not specify the programming language of the code, the code is initially processed through both a rule-based and a neural network model-based language parser to determine the programming language.
The classification error rate of this combined classifier is less than 0.1\% on a dataset of 10 million lines of code. 
Subsequently, the code is comprehensively analyzed by various modules within MPLSandbox. MPLSandbox comprises three core modules: (1) Multi-Programming Language Sandbox Environment, (2) Code Analysis Module, and (3) Information Integration Module.

\textbf{Multi-Programming Language Sandbox Environment.}
Based on the programming language specified for the code, the module first sends the code and unit test samples into the sub-sandbox of the corresponding programming language. 
It then securely compiles and executes the code. 
The sub-sandbox for compilation and execution is a container isolated from the main sandbox environment to prevent potential vulnerabilities and viruses in the code from jeopardizing the external environment during execution. 
Additionally, the sub-sandbox is configured with resource constraints, such as maximum memory limit, maximum execution time, and maximum PIDs limit, to prevent the code from consuming excessive resources that could crash the sandbox environment.
To further ensure the stability of the sandbox during LLM training and deployment, a driver node continuously monitors the state of the sandbox node in real-time and can automatically restart it in case of a crash due to unknown reasons. 
The sandbox environment also monitors and analyzes runtime and resource usage during program execution (detailed in the Code Analysis Module).

Additionally, each programming language sub-sandbox has pre-installed with accordingly widely used dependency libraries. 
Users can also write a configuration file to effortlessly install additional libraries referenced by the code. 
The sandbox environment can also report missing libraries based on compiler feedback, allowing users to easily identify and install the required dependency libraries.

We have predefined eight commonly used programming languages in the sandbox environment, including Python, Java, C++ (C), C\#, Bash, Go, JS, and TS. 
Expanding to additional, unintegrated programming languages is also straightforward. 
Users can create their own sub-sandbox and seamlessly integrate it into the sandbox environment.

\begin{table*}[htbp]
\tiny
  \centering
  \caption{Overview of code analysis tools across various programming languages and tool types. JS and TS denote JavaScript and TypeScript, respectively.
}
  \begin{spacing}{0.8}
    \setlength{\tabcolsep}{1mm}{
        \begin{tabular}{l|cccccccc}
    \toprule
    \toprule
    \textbf{Type} & \textbf{Python} & \textbf{Java} & \textbf{C++ (C)} & \textbf{C\#} & \textbf{Bash} & \textbf{Go} & \textbf{JS} & \textbf{TS} \\
    \midrule
    Basic Information Analysis &   \makecell{ASTPretty \& \\ Pyflowchart}    &   \makecell{Javalang \& \\ Soot}    &   Clang    &  Roslyn     &   -    &  \makecell{GoAst Viewer \& \\ Angr}     &  Joern     & Ts-morph  \\
    \midrule
    Code Smell Analysis &   \makecell{Pylint \& \\ Radon}    &   Pmd    &    CPPCheck   &  StyleCop.Analyzers     &  ShellCheck     &  golangci-lint     &  \makecell{ESLint \& \\ Shkjem}     & \makecell{ESLint \& \\ TSLint} \\
    \midrule
    Code Bug Analysis &  Bandit     & Checkstyle   &  \makecell{PVS-Studio \& \\ CPPCheck}     &  SonarQube     & Shellcheck  &  \makecell{govulncheck \& \\ gosec}     &  NodeJsScan     & Snyk  \\
    \midrule
    Unit Test Analysis &    Coverage   &   Jacoco    &  GCOV     &   Coverlet    & shcov      &  gocov     &  Istanbul & Istanbul  \\
    \midrule
    Code Efficiency Evaluation &   Line\_profile    &  Jprofile   &   Benchmark.NET     & BenchmarkDotNet      & bashprof      &  pprof     & V8 Profiler      & V8 Profiler  \\
    \bottomrule
    \bottomrule
\end{tabular} }%
\end{spacing}
\label{tab:tools}%
\end{table*}%

\textbf{Code Analysis Module.}
\label{sec:model2}
In this module, we integrate numerous traditional analysis tools to offer a comprehensive analysis report for the code from various perspectives.
Additionally, this module can also assess other input information besides the code, including evaluating the coverage of unit tests for the code, aiding researchers in improving the quality of their provided unit test samples.
We categorized the tools into five categories based on their purpose and analysis results: (1) basic information analysis, (2) code smell analysis, (3) code vulnerability analysis, (4) unit test analysis, and (5) code efficiency evaluation.

Specifically, \textbf{(1) Basic information analysis}, including Abstract Syntax Trees (AST) and Control Flow Graphs (CFG), provides LLMs with detailed information on code structure and semantics, and help them better code comprehension. 
This information enhances LLM performance in tasks such as code completion, refactoring, security analysis, and code translation.
\textbf{(2) Code smell analysis} identifies patterns in code that may indicate issues affecting maintainability, readability, and extensibility, such as code complexity, overengineering, and duplicated code. 
It can significantly assist LLMs in various code-related tasks by improving code quality, aiding in code reviews by identifying potential issues, offering refactoring suggestions for cleaner code, and enhancing code understanding through contextual and structural insights.
\textbf{(3) Code bug analysis} is essential in software development for ensuring quality and stability, and it comprises both static and dynamic analysis.
Static analysis detects errors and vulnerabilities without executing code, while dynamic analysis, including fuzz testing, identifies runtime issues. 
These tools enhance LLMs by improving security, aiding debugging, and generating comprehensive documentation, making code more reliable.
\textbf{(4) Unit test analysis} involves evaluating the effectiveness and coverage of user-provided unit tests to ensure code quality and reliability. 
Tools for unit test analysis can help LLMs identify uncovered code lines, generate new test cases, diagnose errors, and offer code quality suggestions, thereby making the development and testing process more efficient and automated.
\textbf{(5) Code efficiency evaluation} involves assessing code performance and resource utilization by analyzing aspects such as time and space complexity, line-leve execution time, and resource usage. 
Code efficiency evaluation can be integrated into LLMs to enhance them in code-related tasks including identifying inefficiencies, pinpointing bottlenecks, providing optimization suggestions, enabling automated improvements, and offering continuous feedback.

In each category of code analysis tools, we integrate commonly used tools for each programming language, as shown in Table~\ref{tab:tools}. 
Additionally, users can effortlessly embed their analysis tools by writing tool templates.
These tools provide more comprehensive information about the code, and their analysis results help LLMs better understand the code. 
The synergistic combination of these tools and LLMs can enhance the performance of LLMs in various code-related tasks. 
We have demonstrated the ease of use and applicability of MPLSandbox by integrating it into several tasks, as detailed in Section~\ref{sec:exp}.

\textbf{Information Integration Module.}
This module can integrate the compiler feedback from the Multi-Programming Language Sandbox Environment, and various code analysis results from the Code Analysis Module into LLM, to improve the quality of generated code and accomplish a range of complex code-related tasks.
Specifically, MPLSandbox includes rich templates that feed the compilation results and analysis outputs to LLMs through prompt learning. 
Users can also easily construct custom prompt templates by combining these results. 
This capability streamlines LLM workflows in various downstream code-related tasks, to reduce development costs. 
For example, users can enable LLMs to generate more diverse and comprehensive unit test examples based on the results of unit test analysis and compiler feedback. 
They can also improve code translation performance by leveraging various structural, semantic, and execution information of the code.

\subsection{Pipeline}

\begin{figure}[htbp]
\centerline{\includegraphics[width=0.48\textwidth]{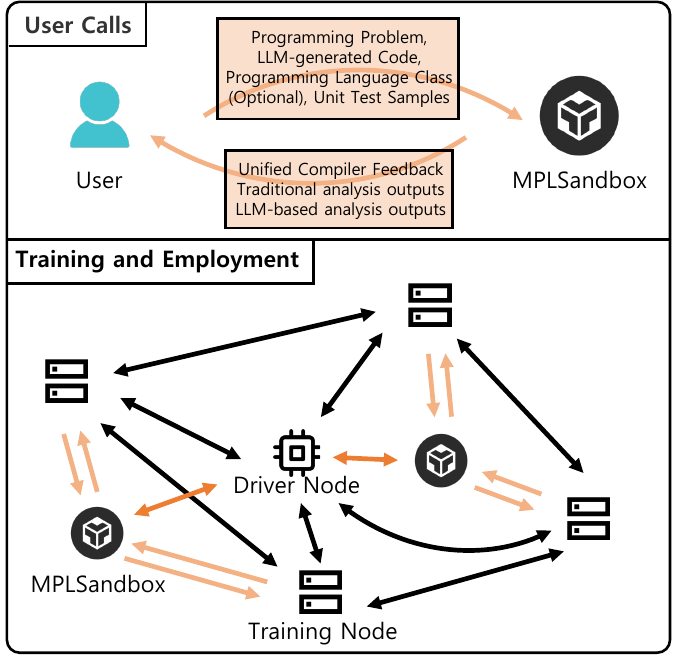}}
\caption{The pipeline of MPLSandbox. 
It can be deployed as a standalone system for users or a few LLMs, or as a distributed system for large-scale LLMs' training and deployment.}
\label{fig:pipe}
\end{figure}

MPLSandbox can be deployed either as a standalone system, serving users or a small number of LLMs, or as a distributed system, serving large-scale training and deployment scenarios. 
The pipeline of MPLSandbox in these two scenarios is illustrated in Figure~\ref{fig:pipe}.
Specifically, users can deploy MPLSandbox on their personal computers or remote servers. 
They can easily invoke MPLSandbox through an IP address and port number to leverage compilers and various analysis tools for comprehensive analysis and evaluation of LLM-generated code cases. 
Additionally, users can integrate MPLSandbox into small-scale LLM training and deployment workflows to enhance their effectiveness.
For instance, MPLSandbox can be integrated into the deployment phase to improve the accuracy and quality of generated code using Best-of-N (BoN). 
It can also be incorporated into the reinforcement learning (RL) training phase by utilizing compiler feedback as an external supervised signal. 
Furthermore, MPLSandbox can optimize workflows for various downstream code-related tasks, as detailed in Appendix~\ref{appendix:casestudy}.

In addition, MPLSandbox can also be easily integrated into large-scale distributed training and deployment environments. Specifically, we deploy multiple sandbox node servers and manage them centrally through a driver node. Sandbox nodes can be custom-assigned to training nodes to provide services.
To prevent sandbox nodes from causing memory and CPU pressure on the training nodes, sandbox nodes and training nodes are deployed separately. 
MPLSandbox streamlines the workflow of large-scale LLM training and deployment, thereby researchers' development time effectively.

\section{Usage}

MPLSandbox is designed to be flexible enough to allow researchers to configure the workflow and integrate their analysis tools, while providing appropriate abstractions to alleviate the concerns of the low-level implementation.
It is ready-to-use and can be easily invoked with just a few lines of code.

\subsection{Initialization}
Firstly, the tool supports initialization as an executor through the following method:

\begin{figure}[htbp]
\vspace{-0.4em}
\leftline{\includegraphics[width=0.65\textwidth]{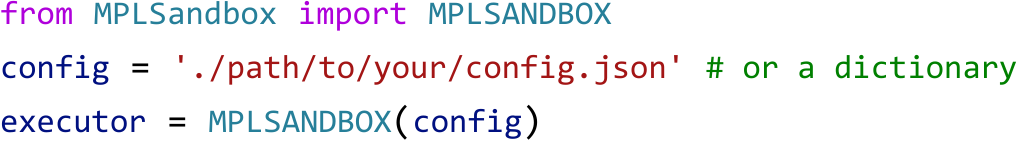}}
\label{fig:initialization}
\vspace{-0.4em}
\end{figure}

where the \lstinline{config.json} file is a configuration file that contains fields such as questions, model-generated code, input and output for unit testing, language types, etc.
Of course, this tool also supports passing configurations directly as a dictionary. Figure~\ref{fig:config} in Appendix~\ref{appendix:usagedetail} shows the detailed configuration for initialization.

\subsection{Run}
Next, this executor can be run through the following method:

\begin{figure}[htbp]
\vspace{-0.4em}
\leftline{\includegraphics[width=0.9\textwidth]{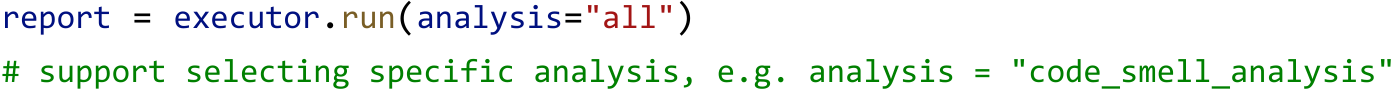}}
\label{fig:run}
\vspace{-0.4em}
\end{figure}

The executor will first call the Code Analysis Module to obtain the 5 types of code analysis information, and then integrate this information through the Information Integration Module before returning it to the user. 
\lstinline{analysis} supports users in defining the code analysis information they need to obtain (default returns all 5 types' information). The Appendix~\ref{appendix:usagedetail} presents detailed case studies demonstrating the five types of analyses performed.

\section{Applications}
\label{sec:exp}


In this section, we introduce code datasets, baseline models, and implementation details of our experiments. 
Then, we present MPLSandbox in three application scenarios to improve the quality of LLM-generated code: (1) acting as a verifier for inference, (2) providing compiler feedback for reinforcement learning, and (3) offering code analysis for self-correction and code optimization.
We also provide more application scenarios and cases of MPLSandbox in a wide range of code-related tasks in Appendix~\ref{appendix:casestudy}.

\subsection{Datasets}

To validate the effectiveness of MPLSandbox, we conduct all experiments on the \textbf{TACO} dataset \cite{li2023taco}, which contains two widely used datasets, \ie APPS \cite{apps} and CodeContests \cite{li2022competition}, and a portion of the newly crawled data from the contest sites.
\textbf{APPS} is a code generation benchmark collected from open-access sites, including Codewars, AtCoder, Kattis, and Codeforces. 
\textbf{CodeContests} is a comprehensive collection of competitive programming problems, designed to facilitate research and development in the fields of code synthesis.
For the APPS part of TACO, we replace the data with APPS+ \cite{dou2024stepcoder}, which is the curated version of APPS.
It excludes instances lacking input, output, or canonical solutions, and eliminates issues such as incomplete code and syntax errors.
The final dataset contains 12,000, 1,000, and 1,000 programming problems for training, validation, and testing, respectively. 
We can train and evaluate LLMs under different programming languages by modifying the system template, as illustrated in Appendix~\ref{appendix:prompts}.

\subsection{Models}
To validate our tool, we integrate it into a wide range of LLMs, including DeepSeek-Coder-Instruct-6.7B \cite{Guo2024DeepSeekCoderWT}, DeepSeek-Coder-V2-Lite-Instruct-16B \cite{zhu2024deepseek}, Qwen2.5-Coder-1.5B-Instruct \cite{qwen2.5}, Qwen2.5-Coder-7B-Instruct \cite{qwen2.5}, Codestral-v0.1-22B \cite{mistral}, Llama-3.1-Instruct-70B \cite{dubey2024llama}, and GPT-4o \cite{openai2023gpt4}.
DeepSeek-Coder-Instruct-6.7B and DeepSeek-Coder-V2-Lite-Instruct-16B are trained by DeepSeek. Qwen2.5-Coder-1.5B and Qwen2.5-Coder-7B are developed by the Qwen group. 
Codestral-v0.1-22B and Llama-3.1-Instruct-70B are developed by Mistral and Meta, respectively. 
GPT-4o is a widely used closed-source LLM developed by OpenAI.
The details of these LLMs are provided in the Appendix~\ref{appendix:models}.

\subsection{Implementation Details}

First, we utilize MPLSandbox as a verifier to verify LLM-generated code at inference time. 
We employed the Best-of-N (BoN) strategy to sample LLM's responses multiple times for the same programming problem, and then evaluate the code using Pass@k metric \cite{chen2021evaluating}. 
In the Pass@1 and Pass@10 settings, the sample temperature is set to 0.2 and 0.8, respectively. 
All inference experiments are conducted on a single node equipped with eight NVIDIA A100-80G GPUs.

Second, we use the compiler feedback provided by MPLSandbox as a supervised signal to optimize LLMs through reinforcement learning. 
We use DeepSeek-Coder-Instruct-6.7B \cite{Guo2024DeepSeekCoderWT} as the foundation LLM.
In each programming language experiment, we modify system prompt templates to enable LLM to solve problems using the according programming language.
All training experiments are conducted on 16 training nodes, totaling 128 NVIDIA A100-80G GPUs, and two MPLSandbox nodes.
The global batch size is set to $512$.
The system prompt template used for constructing multi-programming language instructions is illustrated in Appendix~\ref{appendix:prompts}.
The global training step is set to $8000$, with a 0.1 ratio of warmup.
We report the accuracy point on the test set using the checkpoint at which the model achieves its best performance on the validation set.
The learning rate for the policy model and the critic model is $5e^{-7}$ and $1.5e^{-6}$, respectively.
For each example, we collect a $16$ roll-out code using nucleus sampling. 
The sampling temperature is set to $0.8$, top-p is set to $0.9$, and the maximum output token length is set to $2048$. 
The token-level KL penalty coefficient $\beta$ is set to $0.02$, with a clip value of $0.8$. 
In the decoding phase, the temperature and top-p are set to $0.2$ and $0.95$, respectively.

Finally, through self-reflection and self-correction, we utilize LLMs to correct their generated errors, and improve the quality of codes using the results of analysis tools.
The system prompt templates are shown in Appendix~\ref{appendix:prompts}.
The temperature is set to 0.8.

\subsection{Results}

To showcase our tool, we evaluate it in three application scenarios, to improve the quality of LLM-generated code and help users streamline LLM workflows in various downstream code-related tasks.

\subsubsection{MPLSandbox as a Verifier at inference time}

First, we integrate MPLSandbox into the deployment environment of LLMs to act as a verifier, using the multi-programming language sandbox environment to verify the correctness of generated code at inference time. 
The experimental results are illustrated in Table~\ref{tab:scenario1}. 
The results demonstrate that MPLSandbox can efficiently verify the correctness of model-generated code in multiple programming languages. 
Moreover, the Pass@10 results significantly outperform the Pass@1 results, indicating that MPLSandbox can be stably integrated into various LLMs to provide reliable verification and feedback.

\begin{table*}[htbp]
\small
  \centering
  \caption{
  Performance of LLMs on multi-programming languages integrated with MPLSandbox. JS and TS denote JavaScript and TypeScript, respectively. DS denotes DeepSeek. The results indicate that our tool can provide reliable verification and feedback.
}
  \begin{spacing}{0.6}
    \setlength{\tabcolsep}{0.8mm}{
    \begin{tabular}{r|c|c|cccccccc}
    \toprule
    \toprule
    \multicolumn{1}{l|}{\textbf{Model}} & \textbf{Size} & \textbf{Pass@K} & \textbf{Python} & \textbf{Java} & \textbf{C++ (C)} & \textbf{C\#} & \textbf{Go} & \textbf{Bash} & \textbf{JS} & \textbf{TS} \\
    \midrule
    \multicolumn{1}{l|}{\multirow{2}[0]{*}{Qwen2.5-Coder-Instruct}} & \multirow{2}[0]{*}{1.5B} & K=1   & 2.4\% & 2.8\% & 2.8\% & 0.4\% & 1.1\% & 0.0\% & 0.4\% & 0.4\% \\
          &       & K=10  & 13.9\% & 4.9\% & 8.5\% & 7.3\% & 4.9\% & 4.5\% & 2.4\% & 1.7\% \\
    \midrule
    \multicolumn{1}{l|}{\multirow{2}[0]{*}{Qwen2.5-Coder-Instruct}} & \multirow{2}[0]{*}{7B} & K=1   & 7.0\% & 14.3\% & 11.9\% & 11.5\% & 3.5\% & 4.9\% & 9.1\% & 3.8\% \\
          &       & K=10  & 24.7\% & 23.7\% & 32.1\% & 28.6\% & 23.7\% & 20.6\% & 25.4\% & 17.8\% \\
    \midrule
    \multicolumn{1}{l|}{\multirow{2}[0]{*}{DS-Coder-Instruct}} & \multirow{2}[0]{*}{6.7B} & K=1   & 9.4\% & 10.5\% & 9.1\% & 8.0\% & 3.8\% & 2.4\% & 7.0\% & 3.1\% \\
          &       & K=10  & 23.7\% & 24.7\% & 22.3\% & 25.1\% & 21.6\% & 16.4\% & 21.6\% & 15.3\% \\
    \midrule
    \multicolumn{1}{l|}{\multirow{2}[1]{*}{DS-Coder-V2-Lite-Instruct}} & \multirow{2}[1]{*}{16B} & K=1   & 29.6\% & 26.8\% & 25.1\% & 23.7\% & 10.5\% & 5.6\% & 12.9\% & 8.0\% \\
          &       & K=10  & 50.2\% & 47.7\% & 44.6\% & 42.9\% & 35.5\% & 19.9\% & 39.4\% & 25.1\% \\
    \midrule
    \multicolumn{1}{l|}{\multirow{2}[0]{*}{Codestral-v0.1}} & \multirow{2}[0]{*}{22B} & K=1   & 9.8\% & 21.3\% & 22.0\% & 20.2\% & 12.2\% & 10.1\% & 9.8\% & 7.0\% \\
          &       & K=10  & 34.2\% & 41.8\% & 38.7\% & 41.1\% & 34.8\% & 28.9\% & 34.8\% & 28.6\% \\
    \midrule
    \multicolumn{1}{l|}{\multirow{2}[0]{*}{Llama-3.1-Instruct}} & \multirow{2}[0]{*}{70B} & K=1   &   15.0\% & 17.4\% & 15.7\% & 13.2\% & 6.6\% & 7.4\% & 9.4\% & 6.1\% \\
          &       & K=10  &   38.0\% & 38.3\% & 34.5\% & 35.5\% & 35.5\% & 17.1\% & 33.5\% & 14.6\% \\
    \midrule
    \multicolumn{1}{l|}{\multirow{2}[1]{*}{GPT-4o}} & \multirow{2}[1]{*}{-} & K=1   &    39.3\%   &    47.4\%   &   46.3\%    &   16.0\%    &   43.6\%    &    33.8\%   &    44.6\%   &    40.4\% \\
          &       & K=10  &   52.6\%    &  68.6\%     &  65.9\%    &    47.4\%   &  64.5\%     &    58.2\%   &    66.2\%   &   63.4\% \\
    \bottomrule
    \bottomrule
\end{tabular} }%
\end{spacing}
\label{tab:scenario1}%
\end{table*}%

This integration simplifies various real-world deployment scenarios, including code evaluation, code data production and filtering, and automated testing. 
In practice, we can deploy multiple sandbox nodes where users can verify the correctness of the code and obtain feedback. 
Data production teams can utilize MPLSandbox to filter LLM-generated code in bulk. 
Additionally, MPLSandbox can be integrated into various evaluation environments to provide compiler feedback.

\subsubsection{MPLSandbox for Reinforcement Learning with Compiler Feedback}

\begin{figure}[htbp]
\centerline{\includegraphics[width=0.6\textwidth]{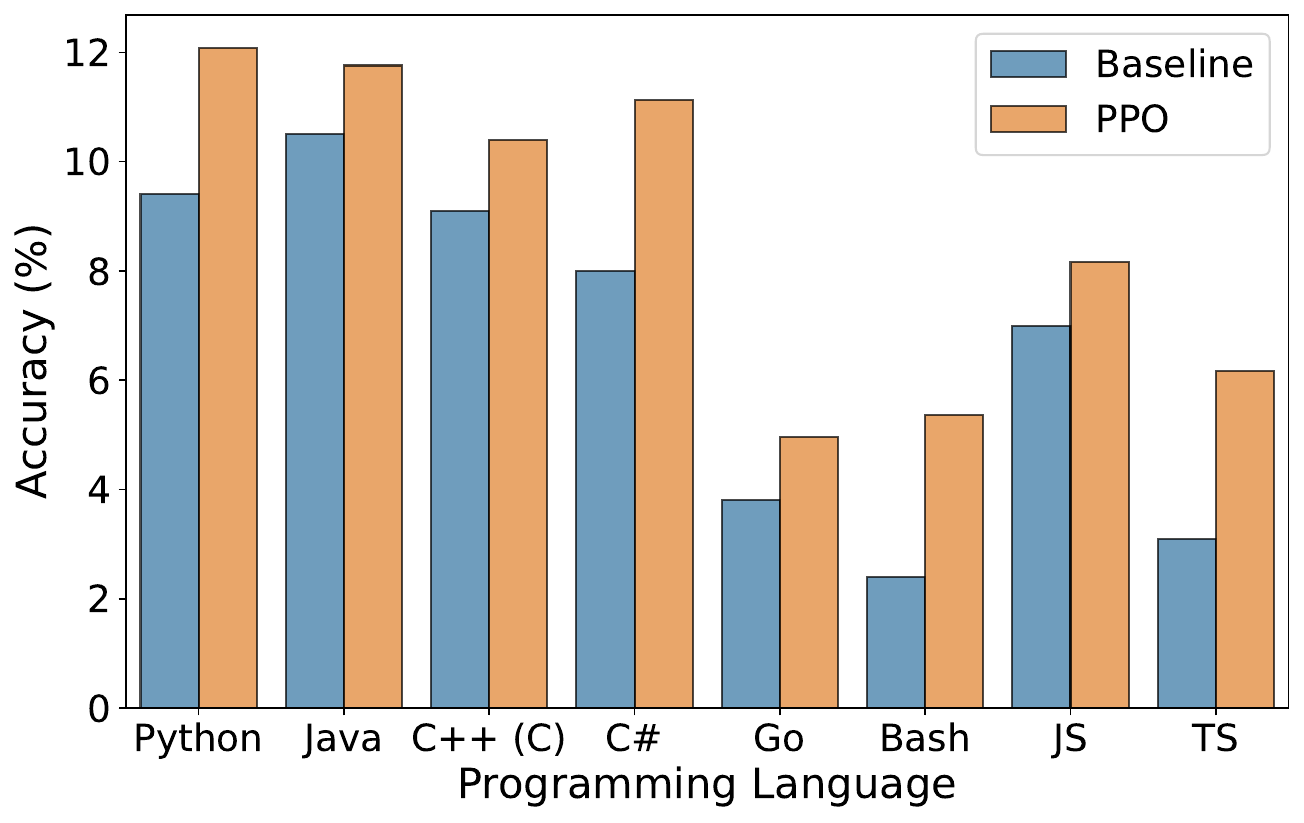}}
\caption{Multi-programming language results of baseline and model trained by PPO. JS and TS denote JavaScript and TypeScript, respectively.
We use DeepSeek-Coder-Instruct \cite{Guo2024DeepSeekCoderWT} as our foundation model and report the Pass@1 results for all programming languages.
Through MPLSandbox, users can easily obtain reliable compiler feedback and effortlessly streamline their LLM training workflow.
}
\label{fig:scenario2}
\end{figure}

We further leverage the compilation feedback provided by MPLSandbox as a feedback signal for reinforcement learning to improve the quality of code generated by LLMs. 
The reinforcement learning algorithm for code generation is detailed in Appendix~\ref{appendix:feedback}.
The experimental results, shown in Figure~\ref{fig:scenario2}, demonstrate that compiler feedback provided by MPLSandbox is effective and stable, significantly enhancing LLM performance in code generation.
Users can access MPLSandbox with just a few lines of code to obtain compiler feedback for their code. 
This compiler feedback is then standardized into unified signals regardless of the programming language. 
Users can also customize the rules to define the numericalization method.

In summary, MPLSandbox allows users to avoid focusing on trivial tasks such as isolating and building a multi-programming language execution environment. 
By simply invoking MPLSandbox, users can obtain robust compiler feedback, enabling them to devote more time to developing and optimizing their training algorithms.

\subsubsection{MPLSandbox for Self-Correction and Code Optimization}

Self-correcting and optimizing codes generated by LLMs is crucial but often complicated and tedious. 
This process requires providing LLMs with detailed information about code errors and various characteristics of the code, such as complexity, execution efficiency of each line, and adherence to coding standards. 
Achieving this involves developing a sandbox environment, utilizing various analysis tools, and integrating their results.
With MPLSandbox, users can effortlessly compile and analyze LLM-generated code. 
You can also easily customize the system prompt template to combine these results in various ways.

As a demonstration, we utilize compiler feedback to enable GPT-4o to correct erroneous code and code smell analysis results to refine the correct code, respectively.
The system prompts are shown in Appendix~\ref{appendix:prompts}.

\begin{table*}[htbp]
\small
  \centering
  \caption{Improvement results after self-correction and self-refinement show significant enhancements 
}
  \begin{spacing}{0.6}
    \setlength{\tabcolsep}{0.8mm}{
    \begin{tabular}{l|cccccccc}
    \toprule
    \toprule
    \textbf{Metric} & \textbf{Python} & \textbf{Java} & \textbf{C++(C)} & \textbf{C\#} & \textbf{Go} & \textbf{Bash} & \textbf{JavaScript} & \textbf{TypeScript} \\
    \midrule
    Pass@1    & +3.7\%   & +4.9\%   & +2.7\%  & +6.5\%  & +5.0\%   & +4.8\%   & +4.1\%   & +3.1\%  \\
    Avg num of comment lines & +8.9  & +7.0  & +7.2  & +5.2  & +5.6  & +4.3  & +7.9  & +6.3  \\
    \bottomrule
    \bottomrule
\end{tabular} }%
\end{spacing}
\label{tab:scenario3}%
\end{table*}%

The experimental results, as shown in Table~\ref{tab:scenario3}, show that by using MPLSandbox, GPT-4o can effectively and efficiently correct and refine its generated code.
Through self-correction and self-refinement, GPT-4 can solve more programming problems, and the generated code includes more comments. 
Moreover, the generated code exhibits lower complexity and is more compliant with programming specifications, as detailed in Appendix~\ref{appendix:ins}.
This will help developers understand and maintain the programs generated by LLMs.
The results demonstrate that our tool enables users to implement LLM self-correction and optimization easily. 
This can assist users in serving downstream tasks more conveniently and efficiently, such as improving the quality of LLM-generated code at inference time and producing reflection data for training.

We also provide more application scenarios and cases of MPLSandbox in a wide range of code-related tasks in Appendix~\ref{appendix:casestudy}, including unit test generation, vulnerability localization, and code translation.
These cases and results indicate that MPLSandbox is effective for most workflows, significantly reducing the user's development effort.

\section{Related Work}
In this section, we review existing works on LLMs used for code-related tasks, and introduce existing code analysis tools.

\subsection{Large Language Models for Code}
Recently, the advancement of LLMs \cite{jiang2023mistral, bai2023qwen, deepseek-llm, gpt35, openai2023gpt4, anthropic2024claude, touvron2023llama, llama2, llama3, lu2021codexglue, allal2023santacoder, chen2021evaluating, abdin2024phi} has significantly propelled the field of software engineering \cite{dou2024s, jin2024llms, xu2024large, nam2024using}.
For instance, in code generation and program repair, state-of-the-art approaches improve the correctness and quality of LLM-generated codes by learning from compiler feedback.
Specifically, researchers integrate compiler information into prompt templates to improve the performance of LLMs on these code-related tasks \cite{ren2024reflectioncoder, jiang2024training, le2023codechain, shin2023prompt, denny2023conversing}.
Some work also transforms compiler information into feedback signals to optimize LLMs to enhance their performance \cite{dou2024stepcoder, yu2023mathcal, le2022coderl, liu2023rltf, shojaee2023execution}.
For security, stability, reliability, and providing robust monitoring capability, these compilation and execution processes are needed in an isolated sandbox \cite{garfinkel2003virtual, Guo2024DeepSeekCoderWT, liang2003isolated}.
However, the development of open-source sandboxes is still in its early stages.
Meanwhile, existing sandboxes developing for LLM-generated code mostly focus on a single programming language, such as Python \cite{engelberth2012pybox, Terrarium, promptfoo} or a few programming languages \cite{LLMSandbox}, which lacks numerous commonly used dependency libraries.
The absence of an easy-to-use sandbox necessitates that researchers spend considerable time on installing environments and dependency packages, as well as constructing a distributed sandbox to support multiple programming languages.

Some researchers also combine compiler feedback with other analysis results of generated codes to enhance LLM's performance on more code-related tasks \cite{lin2020software, korel1990automated, lowry1969object, roziere2020unsupervised, ahmad2020transformer, antoy1990specification}.
For instance, some work focuses on developing LLM-based software vulnerability detection approaches by using compiler feedback and traditional code features \cite{lu2024grace, du2024generalization}.
Some work also aims to enhance LLM's performance in generating more diverse unit test samples (\ie software test generation) \cite{gu2024testart, ryan2024code, chen2024chatunitest}, generating more comprehensive code specification (\ie specification generation) \cite{ma2024specgen, zheng2023towards, jin2024llms}, optimizing the code snippets (\ie code efficiency optimization) \cite{gao2024search, du2024mercury}, summarizing the code with natural language (\ie code summary) \cite{virk2024enhancing, kumar2024code, nam2024using}, and translating code from one programming language to another programming language (\ie code translation) \cite{pan2024lost, yin2024rectifier, bhattarai2024enhancing}.
The application of LLMs to code-related tasks necessitates leveraging a vast array of traditional analysis tools from the field of software engineering. 
Researchers will expend much time and effort on trivial work such as constructing the environment and resolving versioning and dependency issues. 
An out-of-the-box framework that integrates a multilingual isolated compilation and execution sandbox with multilingual analysis tools remains unexplored.

\subsection{Program Analysis Tools}

The analysis of structural information, intermediate variable flow during execution, and resource usage can help researchers comprehensively understand and improve the quality of code \cite{wang2024python, ryan2024code}. 
For instance, various tools are employed to obtain code structure information and code smells, such as abstract syntax trees (AST) (\eg Python's ASTPretty \cite{astpretty} and Java's Javalang \cite{javalang}), control flow graphs (CFG) (\eg C++'s Clang \cite{clang} and JS's Joern \cite{joern}), complexity measurement tools (\eg Python's Radon \cite{radon} and Java's Pmd \cite{pmd}), and unit test coverage analysis tools such as Coverage \cite{coverage}.
Fuzzing test tools employ techniques such as coverage-based feedback-driven testing and fault injection to discover security vulnerabilities in programs \cite{xia2024fuzz4all}. 
Additionally, tools that evaluate code efficiency, such as performance profilers (\eg Line\_profile \cite{lineprofiler} and Jprofile \cite{jprofiler}), help researchers analyze, optimize, and maintain high-quality programs. 

These program analysis tools can also be utilized by LLMs to enhance various downstream code-related tasks \cite{du2024generalization, cheng2024llm}.
We have predefined five categories of tools and integrated open-source mainstream tools for each programming language, redirecting the output of these tools to text. 
Additionally, more options for program analysis tools are available.
We also design templates that allow users to integrate their own tools with ease.
MPLSandbox can unify results from compiler feedback and a variety of powerful analysis tools to streamline LLM workflows in a wide range of code-related tasks.

\section{Conclusion}
In this paper, we introduce MPLSandbox, an out-of-the-box multi-programming language sandbox designed to provide unified compiler feedback and comprehensive analysis using both traditional-based and LLM-based methods for LLM-generated code. 
MPLSandbox not only can be effortlessly used by researchers to thoroughly analyze LLM-generated codes, but also can be integrated into the training and deployment phases to improve the correctness and quality of generated codes. 
Additionally, MPLSandbox can enhance the performance of LLMs across a wide range of code-related tasks through the flexible combination of various analysis tools within the sandbox.
The goal of MPLSandbox is to support and advance further research in LLMs for software engineering by simplifying the complexity of training and employing LLMs in code-related tasks.

\bibliographystyle{ACM-Reference-Format}
\bibliography{Sandbox}

\appendix

\newpage

\section{Case Study on Usage}
\label{appendix:usagedetail}

\begin{figure}[htbp]
\vspace{-0.5em}
\centerline{\includegraphics[width=0.7\textwidth]{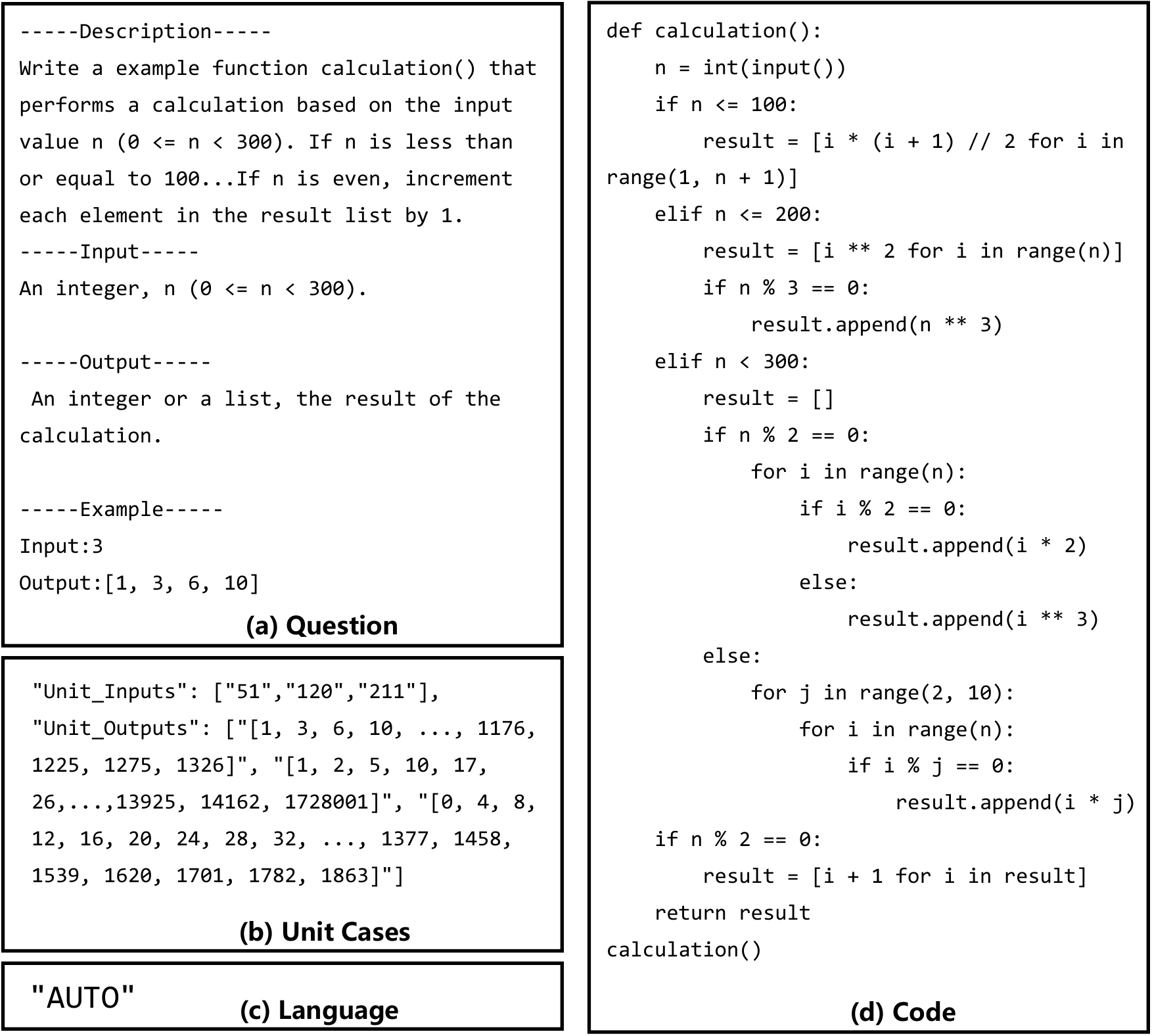}}
\caption{The Form of Configuration}
\label{fig:config}
\vspace{-0.5em}
\end{figure}

This section demonstrates the usage of various analysis function of MPLSandbox through a configuration example shown in Figure~\ref{fig:config}, where: 

\begin{itemize}
    \item \lstinline{Question} field presents the code generation problem posed to the LLM, requiring a function named \lstinline{calculation(n)} that performs complex calculations based on different input values of \lstinline{n}. The problem description provides the input requirements (an integer, ranging from 0 to 299), output requirements (an integer or a list, depending on the result of the calculation), and an example (when the input is 3, the output should be [1, 3, 6, 10]).
    \item \lstinline{Code} field is the code provided by the LLM in response to the \lstinline{Question} field.
    \item \lstinline{Unit Cases} field provides 3 unit test cases, which includes two sub-fields: \lstinline{Unit Inputs} and \lstinline{Unit Outputs}. \lstinline{Unit Inputs} shows three test inputs provided: "51", "120", "211". \lstinline{Unit Outputs} shows the corresponding expected output results provided.
    \item \lstinline{Language} field specifies the language of the code, which is set to "AUTO" here, meaning the code language is automatically detected.
\end{itemize}

Moreover, it still supports specifying docker client instance, docker image, and dockerfile for building custom Docker images. For more details about parameter configuration, please move to: https://github.com/Ablustrund/MPLSandbox 

After initializing MPLSandbox as an executor using the configuration, it can be run by calling the \lstinline{run()} method. MPLSandbox analyzes the code across five levels by calling the Code Analysis Module and integrates the analysis results through the Information Integration Module. Users can choose the required related information according to their needs. The following sections will conduct corresponding case studies on the example code in the configuration.

\subsection{Basic Information Analysis}

\begin{figure}[htbp]
\centerline{\includegraphics[width=1\textwidth]{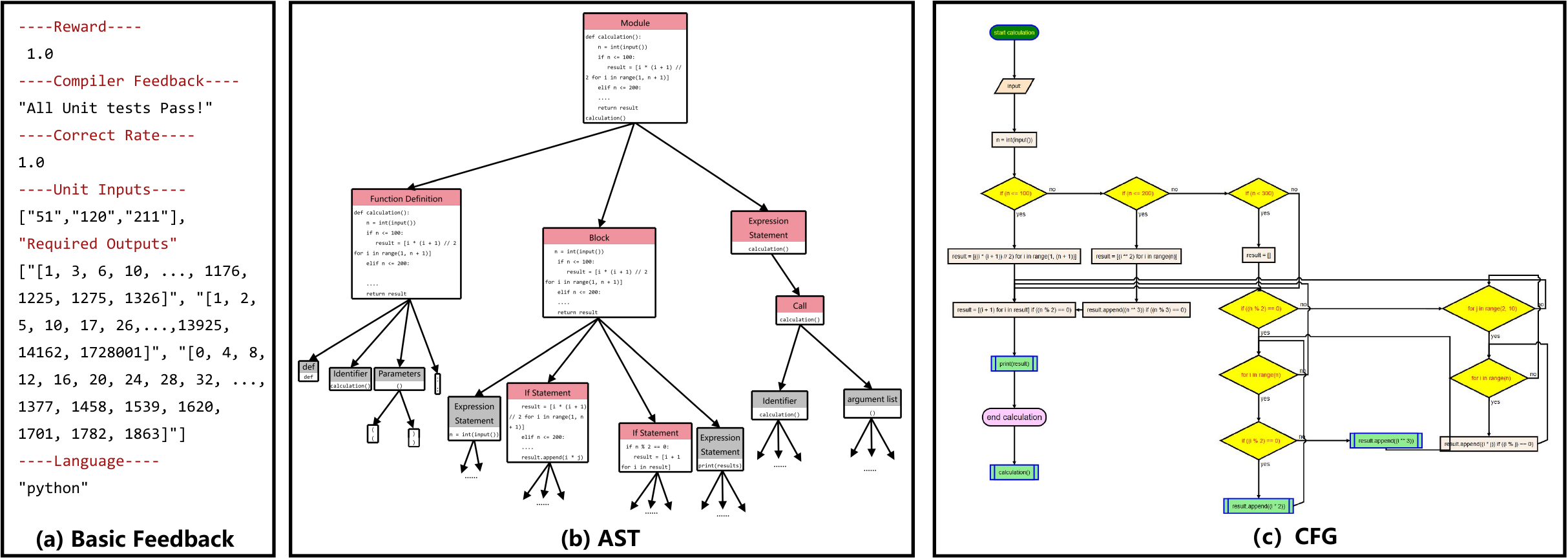}}
\caption{The Report of Code Basic Analysis}
\label{fig:code_basic_analysis}
\end{figure}

As shown in Figure~\ref{fig:code_basic_analysis}, Code Basic Analysis returns a Basic Feedback along with Abstract Syntax Tree (AST) and Control Flow Graph (CFG). The basic feedback includes fields such as \lstinline{Reward}, \lstinline{Compiler Feedback}, \lstinline{Correct Rate}, \lstinline{Unit Inputs}, \lstinline{Required Outputs} and \lstinline{Language}. From the compiler feedback, it can be seen that the code has successfully passed all unit tests, achieving a correct rate of 1.0 and a reward of 1.0. These values can serve as signals for downstream analysis or training tasks. 

Additionally, MPLSandbox parses the code into the forms of an AST and a CFG. The AST presents the syntactic structure of the code in a tree diagram, where each node represents a syntactic element in the code, such as function definitions, assignment operations, conditional judgments, and loops. This structure helps to understand the logic and hierarchical relationships of the code, facilitating code optimization and error detection. The CFG, on the other hand, graphically displays the execution paths and decision points of the code, including basic blocks (representing a series of consecutive instructions) and edges (indicating the direction of control flow), which helps to reveal the execution order of the program and potential branching conditions. The CFG is beneficial for identifying loop dependencies, potential performance bottlenecks, and logical errors in the program.

\subsection{Code Smell Analysis \& Code Bug Analysis}

\begin{figure}[htbp]
\centerline{\includegraphics[width=0.8\textwidth]{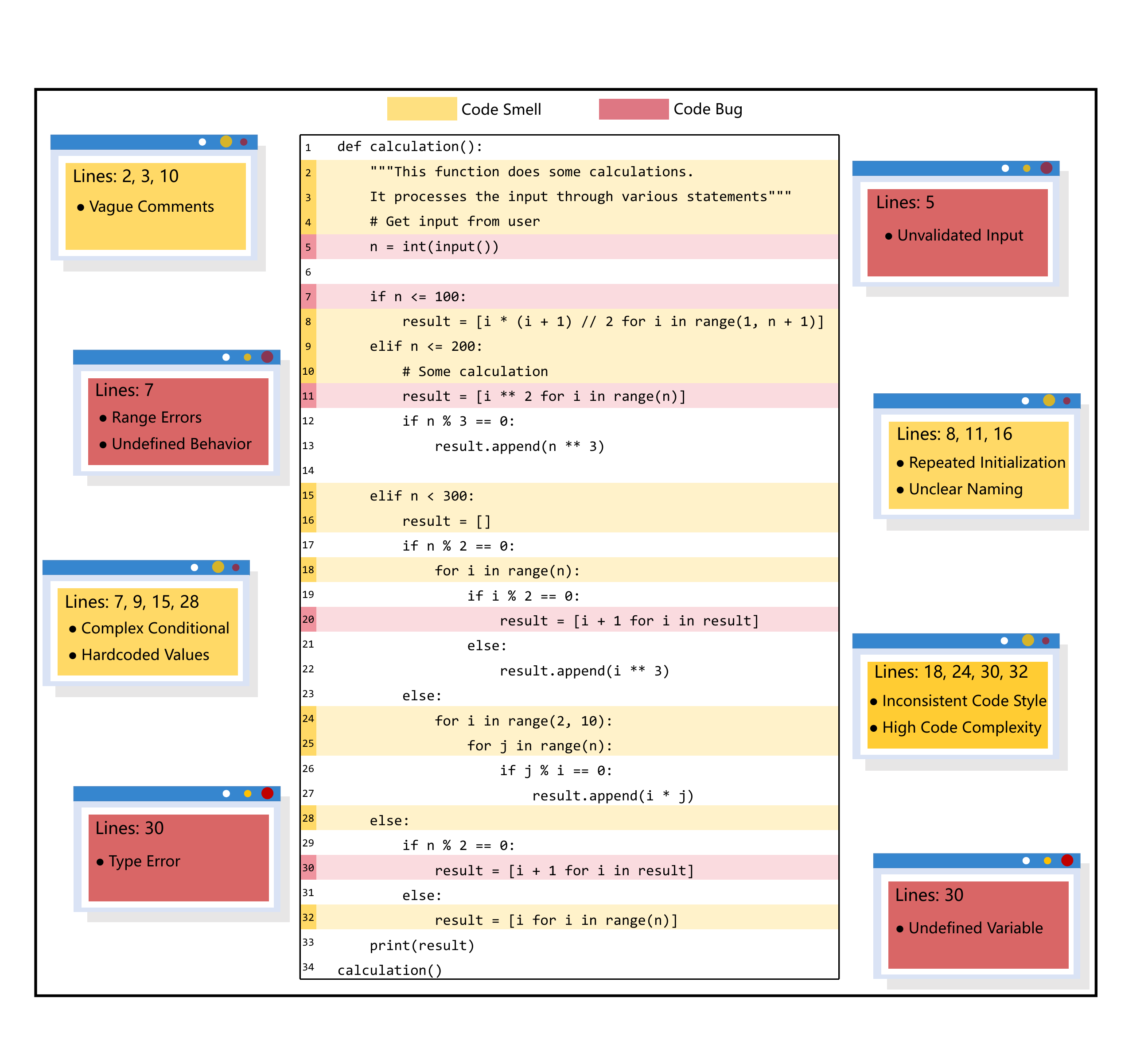}}
\caption{The Report of Code Smell Analysis and Code Bug Analysis.}
\label{fig:code_smell&code_bug}
\end{figure}

Code Smell Analysis and Code Bug Analysis identify potential issues or vulnerabilities within the code, reporting the specific line numbers as well as the categories of smells or bugs. To better demonstrate Code Smell Analysis and Code Bug Analysis, we have artificially introduced some code smells and vulnerabilities into the code. The yellow warning boxes in Figure~\ref{fig:code_smell&code_bug} represent the parts of the code where MPLSandbox has detected code smells, which include:
\begin{itemize}
\item Line 2, 3, and 10 exhibit the Vague Comments smell due to the docstring and comments being filled with vague and abstract descriptions, not providing concrete information to help understand the actual functionality of the function.
\item The Complex Conditional smell on lines 7, 9, 15, 28 is due to the use of multiple if-elif statements within the function, making the conditional logic complex and difficult to trace. The Hardcoded Values smell refers to the values 100, 200, and 300 being hardcoded within the function, requiring changes in multiple places if these values need to be modified.
\item The Repeated Initialization smell on lines 8, 11, 16 is due to the variable \lstinline{result} being initialized in multiple branches, violating the DRY (Don't Repeat Yourself) principle. Unclear Naming refers to the variable \lstinline{result} not clearly expressing its meaning and a more descriptive name might be better.
\item The Inconsistent Code Style smell on lines 18, 24, 30, 32 refers to the inconsistent use of variables \lstinline{i} and \lstinline{j} in for loops across different branches, which may cause confusion. The High Code Complexity smell indicates that the function contains multiple conditional branches and nested loops, making it difficult to understand and maintain the code.
\end{itemize}

\begin{figure}[htbp]
\centerline{\includegraphics[width=0.7\textwidth]{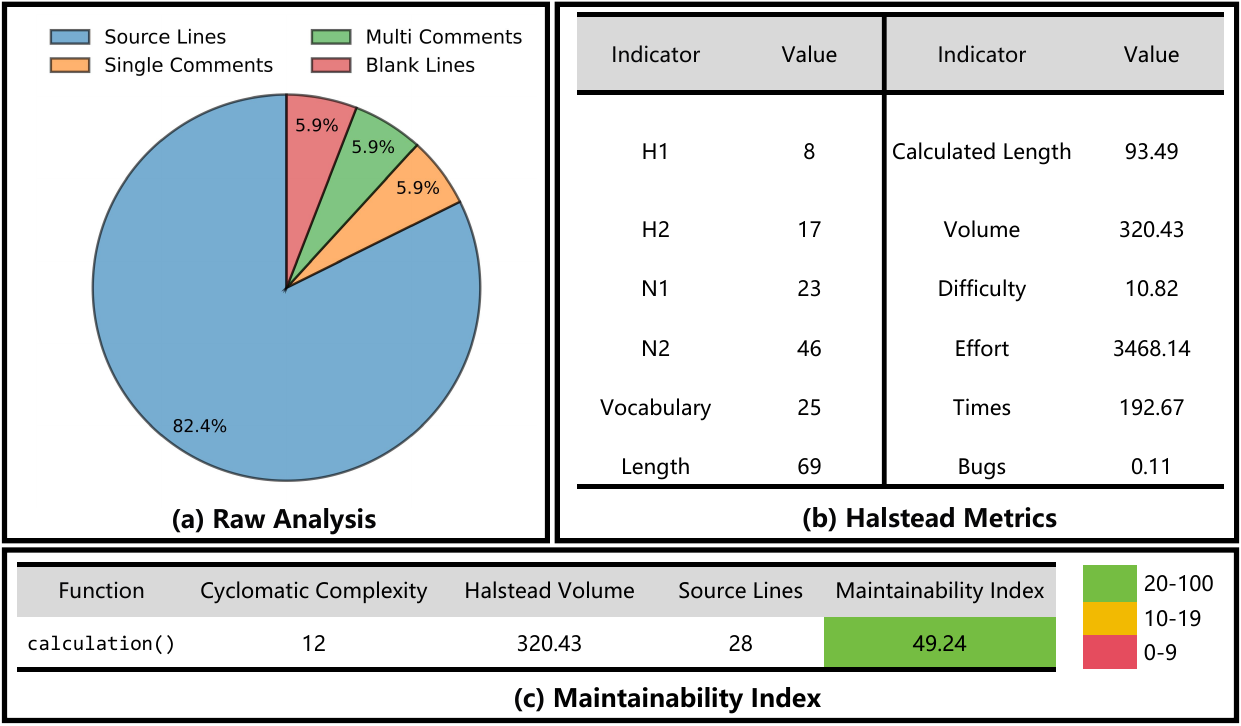}}
\caption{The Sub Report of Code Maintainability Analysis}
\label{fig:code_maintainability_analysis}
\end{figure}

It is worth noting that the bad smell determines the overall maintainability of the code, which is often difficult to quantify. Therefore, Code Smell Analysis provides a sub-report specifically for analyzing the maintainability of the code. As shown in Figure~\ref{fig:code_maintainability_analysis}, the report is divided into three parts: Raw Analysis, Halstead Metrics, and Maintainability Index. The raw analysis provides the distribution of source code, comments, multiline comments, and blank lines, showing a preliminary state of the code. Halstead Metrics offer various indicators of the Halstead volume to quantify the complexity of the code. The Maintainability Index, on the other hand, is a comprehensive calculation of the overall Maintainability Index based on the number of lines of source code, the volume indicator in the Halstead volume, and the cyclomatic complexity of the code, providing a comprehensive assessment of the code's complexity. It's important to note that the Maintainability Index in the report is a standardized result, with its range limited to 0-100. This threshold can be broken down into 0-9 (red), indicating that the code is difficult to maintain. Code within this range may have many issues and requires significant effort to improve its maintainability. 10-19 (yellow) indicates that the code's maintainability is moderate. Although not the worst, code within this range may still require some improvements to enhance its maintainability. 20-100 (green), which is the range where the code falls, indicates that the code has a good structure and clear coding style, making it easy to maintain.

The red warning boxes in Figure~\ref{fig:code_smell&code_bug} represent the parts of the code where MPLSandbox has detected code bugs, which include:
\begin{itemize}
\item The Unvalidated Input bug on line 5 is due to user input being used directly in calculations without any validation or restriction.
\item The Range Errors bug on line 7 occurs when \lstinline{n=0}, causing \lstinline{range(1, n + 1)}  to raise a ValueError. The Undefined Behavior bug is when \lstinline{n} is negative, the behavior of \lstinline{range(n)} is undefined.
\item The Type Error bug on line 20 occurs when the variable \lstinline{result} is empty, executing \lstinline{result = [i + 1 for i in result]} will raise a TypeError.
\item The occurrence of an Undefined Variable on line 30 is due to the fact that the variable \lstinline{result} was not initialized within this conditional branch. Consequently, the \lstinline{result = [i + 1 for i in result]} is invalid code.
\end{itemize}

\subsection{Unit Test Analysis}

\begin{figure}[htbp]
\centerline{\includegraphics[width=0.7\textwidth]{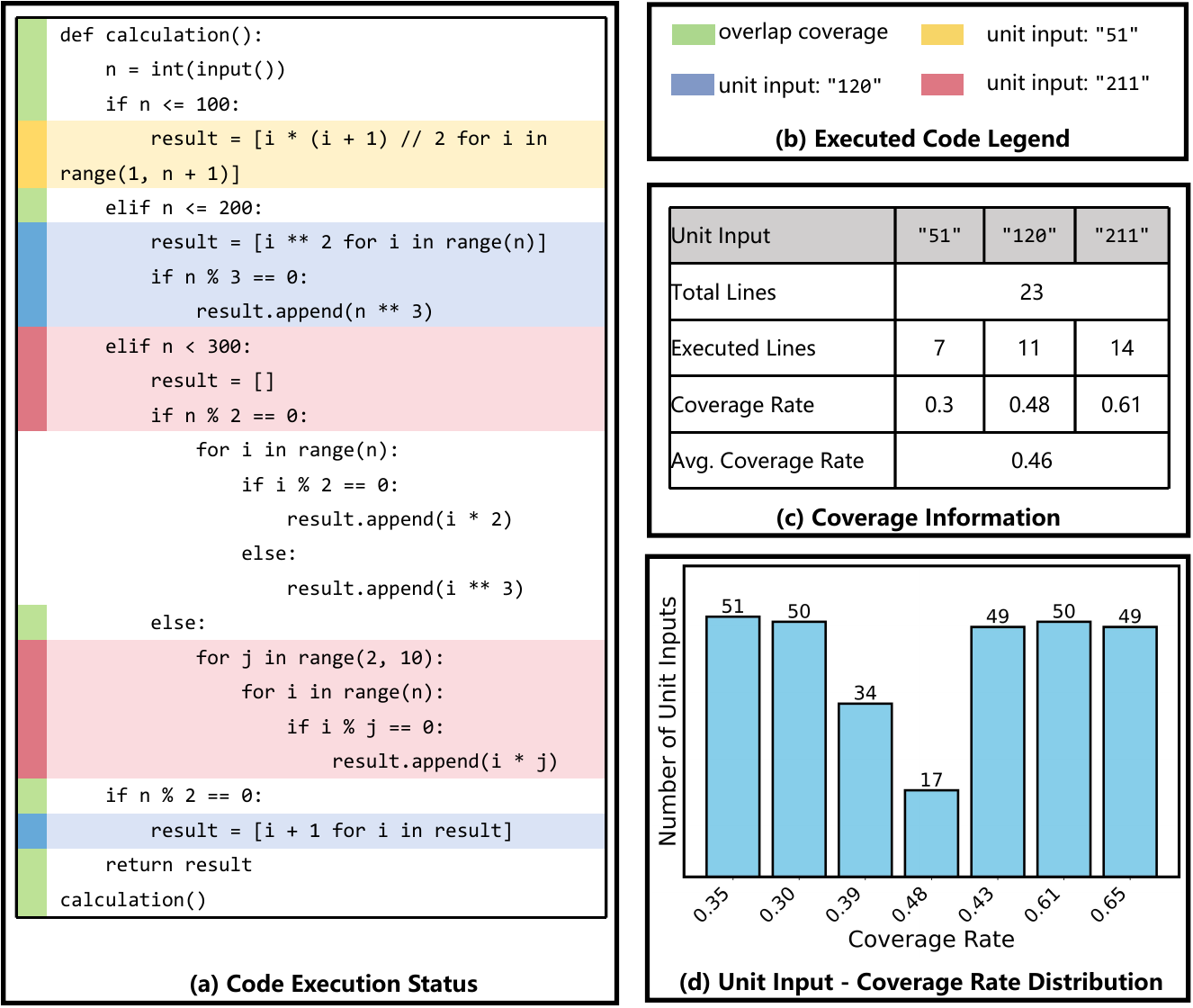}}
\caption{The Report of Unit Test Analysis}
\label{fig:unit_test_analysis}
\end{figure}

Unit Test Analysis returns a comprehensive coverage report for the given unit tests. As shown in Figure~\ref{fig:unit_test_analysis}, green lines represent the overlapping parts of the executed lines for different unit inputs, while yellow, blue, and red lines represent the non-overlapping parts of the executed lines for the test cases "51", "120", and "210", respectively. For the unit input "51", 7 lines of code were executed, with a coverage rate of 0.3. For the unit input "120", 11 lines of code were executed, with a coverage rate of 0.48. For the unit input "211", 14 lines of code were executed, with a coverage rate of 0.61. For a total of 23 lines of code, the overall average coverage rate is 0.46. This indicates that the current test cases do not fully cover the code paths.

Furthermore, Unit Test Analysis has conducted a complete coverage statistics for all test inputs within the given range. It can be observed that within the range of unit input 0 <= n < 300, this set of code has resulted in 7 different coverage possibilities, with the highest being 0.65 and the lowest being 0.35. The distribution of unit inputs across various coverage rates is relatively even. It is evident that after iterating through all possible test inputs, the code coverage remains at a relatively low level, suggesting that the logical framework of the code itself still has significant room for improvement.

\subsection{Code Efficiency Evaluation}

\begin{figure}[htbp]
\centerline{\includegraphics[width=0.96\textwidth]{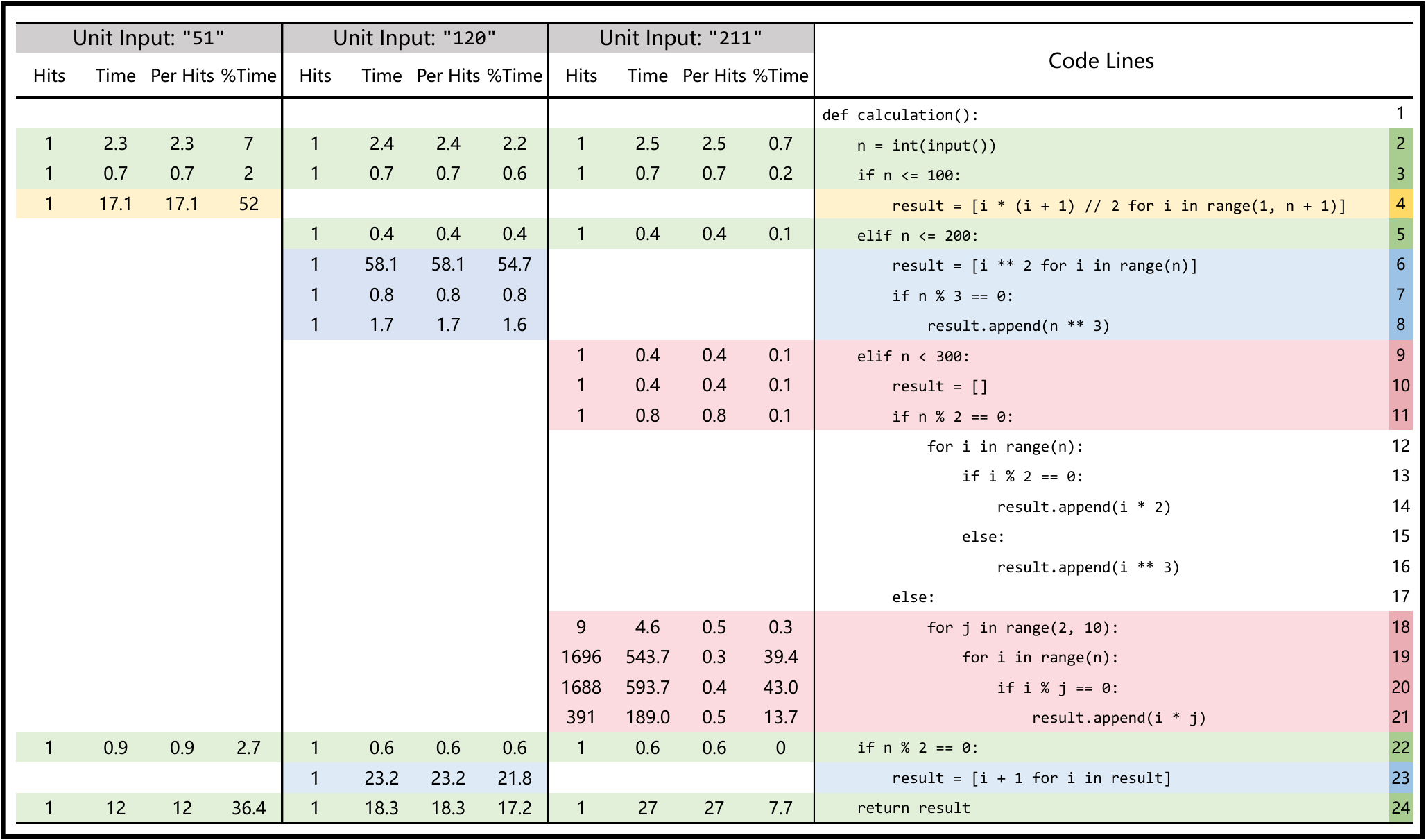}}
\caption{The Report of Code Efficiency Evaluation}
\label{fig:code_efficient_eval}
\end{figure}

Code Efficient Evaluation provides an analysis of code execution efficiency for different test cases. Figure~\ref{fig:code_efficient_eval} reports the Hits (the number of times a code line is executed), Time (the total execution time of the code line in milliseconds), Per Hits (the average time required for each execution of the code line in milliseconds), and \%Time (the percentage of the total execution time taken by the execution time of the code line). 

As shown in Figure~\ref{fig:code_efficient_eval}, code lines 2, 3, 5, 22 and 24 have common execution records under different test inputs, with some code lines taking a longer execution time under specific inputs. For example, code line 6 takes 58.1 milliseconds to execute under the input "120" because in this case, line 6 is a loop that iterates 120 times. Code line 23 takes 33.2 milliseconds to execute under the input "210" because this line of code contains a loop that iterates based on the variable \lstinline{result}, which is strongly related to the input 210. Code lines 12, 13, and 14 have a large number of executions under the input "210" (211 times, 210 times, and 105 times, respectively), because this part involves the processing of a large range loop. Therefore, these perceptions of code line execution efficiency undoubtedly provide very important basis for further performance optimization.

\section{Prompts in Detail}
\label{appendix:prompts}
The system prompt for multi-programming language is as follows:

\begin{tcolorbox}[colback=black!5!white, colframe=black!75!black]
\textit{You are a master programmer. Now given a programming problem, you need to analyze it carefully and answer this programming problem directly in \textbf{\{lang\}} programming language.} \\
\textit{Note that generated code will accept a unit test case through standard input (stdin) from the terminal. And after printing the output result, the program will terminate} \\
\textit{\#\#\# Prompt:} \\
\textit{[Programming problem]} \\
\textit{\textbf{\{programming problem\}}} \\
\textit{Solve this problem directly in \textbf{\{lang\}}. The code needs to be surrounded by backquotes (i.e., ```code```)} \\
\textit{\#\#\# Response:} 
\end{tcolorbox}


The system prompts for correcting and refining code are as follows:

\begin{tcolorbox}[colback=black!5!white, colframe=black!75!black]
\textit{You are a master of program correction. Now given a programming problem, an incorrect solution to this problem, and the compiler error message. You need to carefully analyze and correct the incorrect code} \\
\textit{\#\#\# Prompt:} \\
\textit{[Programming problem]} \\
\textit{\textbf{\{programming problem\}}} \\
\textit{[Incorrect code]} \\
\textit{\textbf{\{incorrect code\}}} \\
\textit{[Compiler error message]} \\
\textit{\textbf{\{error message\}}} \\ 
\textit{Please correct this problem, and the new code needs to be surrounded by backquotes (i.e., ```code```)} \\
\textit{\#\#\# Response:} 
\end{tcolorbox}

\begin{tcolorbox}[colback=black!5!white, colframe=black!75!black]
\textit{You are a master of program refinement. Now given a programming problem, the correct solution, and some suggestions for improvement including readability, complexity and code Specification. You need to refine the code according to these suggestions} \\
\textit{\#\#\# Prompt:} \\
\textit{[Programming problem]} \\
\textit{\textbf{\{programming problem\}}} \\
\textit{[Original code]} \\
\textit{\textbf{\{code\}}} \\
\textit{[Suggestions through code smell]} \\
\textit{\textbf{\{analysis results\}}} \# including code style and specifications (\eg long method, long parameter list and tight coupling) \\
\textit{Please first summarize recommendations for refinement from these analysis results, and then refine this code. The new code needs to be surrounded by backquotes (i.e., ```code```)} \\
\textit{\#\#\# Response:} 
\end{tcolorbox}

\section{Experiments Setup in Detail}
\label{appendix:exp}

\subsection{Models}
\label{appendix:models}

\begin{itemize}

\item \textbf{DeepSeek-Coder-Instruct-6.7B} \cite{Guo2024DeepSeekCoderWT}, developed by DeepSeek AI, demonstrates state-of-the-art performance among open-source code models across various programming languages. The training corpus for these models comprises an impressive 2 trillion tokens, combining code and natural language texts. Each model is trained to utilize a window size of 16K.

\item \textbf{DeepSeek-Coder-V2-Lite-Instruct-16B} \cite{zhu2024deepseek} is an open-source Mixture-of-Experts (MoE) code language model that achieves performance comparable to GPT-4 Turbo in code-specific tasks. It has 2.4 billion active parameters. The model is further pre-trained from an intermediate checkpoint of DeepSeek-V2 with an additional 6 trillion tokens. DeepSeek-Coder-V2 expands its support for programming languages from 86 to 338, while extending the context length from 16K to 128K.

\item \textbf{Qwen2.5-Coder-1.5B-Instruct} \cite{qwen2.5} and \textbf{Qwen2.5-Coder-7B-Instruct} \cite{qwen2.5} are members of the Qwen2.5 series. They are built on the Qwen2.5 architecture and have been further pre-trained on an extensive dataset exceeding 5.5 trillion tokens, which includes source code, text-code grounding data, and synthetic data. These two LLMs support up to 128K tokens of context and cover 92 programming languages.

\item \textbf{Codestral-v0.1-22B} \cite{mistral} is trained by Mistral AI. This model is trained on a diverse dataset encompassing over 80 programming languages, including popular ones like Python, Java, C, C++, JavaScript, and Bash. It also excels in more specialized languages such as Swift and Fortran. This model supports up to 32K context window.

\item \textbf{Llama-3.1-Instruct-70B} \cite{dubey2024llama} is developed by Meta AI. It is pre-trained and instruction-tuned generative models that handle text input and output. The Llama 3.1 instruction-tuned models are specifically optimized for multilingual dialogue applications, surpassing many open-source and proprietary chat models on standard industry benchmarks. The context window of this model is 128K.

\item \textbf{GPT-4o} \cite{openai2023gpt4} is an advanced iteration of OpenAI's language model, designed to offer more refined and contextually aware responses.
GPT-4o is exceptionally versatile, and capable of excelling in a wide range of applications from creative writing to technical problem-solving.
The context window and maximum output tokens are 128K and 16,384, respectively.

\end{itemize}

\subsection{Reinforcement Learning from Compiler Feedback}
\label{appendix:feedback}

Following prior work \cite{dou2024stepcoder, liu2023rltf} in code generation, we use Proximal Policy Optimization (PPO) algorithm \cite{schulman2017proximal} to optimize the policy model $\pi_\theta$ with parameter $\theta$.
Consider a programming problem $x$, the policy model generates the response $\hat{y}$ according to $x$.
The reward model provides the reward $r$ according to the correctness of the code snippet $\tau$ extracted from response $\hat{y}$, where we use the same setting as previous approaches \cite{le2022coderl, shojaee2023execution, dou2024stepcoder} as follows:
\begin{align}
\label{eq:reward}
r(x, \tau) = 
\begin{cases}
~+1,~ \text{if $\tau$ passed all unit tests}\\
-0.3,~ \text{if $\tau$ failed any unit test}\\
-0.6,~ \text{if $\tau$ happened runtime error}\\
~-1,~ \text{if $\tau$ happened compile error}.
\end{cases}
\end{align}

The policy model $\pi_\theta$ can be optimized by maximizing the objection function as follows:
\begin{align}
\label{eq:vanilla-rl}
\small
    \text{Objective}(\theta) = E_{(x, \hat{y}) \sim D_{\pi_\theta}} [r(x, \tau) - \beta \log (\pi_\theta (\hat{y} | x) / \pi^{\text{ref}} (\hat{y} | x)) ]
\end{align}
where $\pi^{\text{ref}}$ is the reference model in PPO, which is initialized with the parameters of the initial policy model and kept frozen throughout the training process.

\section{Instances}
\label{appendix:ins}

\begin{figure}[htbp]
\centerline{\includegraphics[width=1\textwidth]{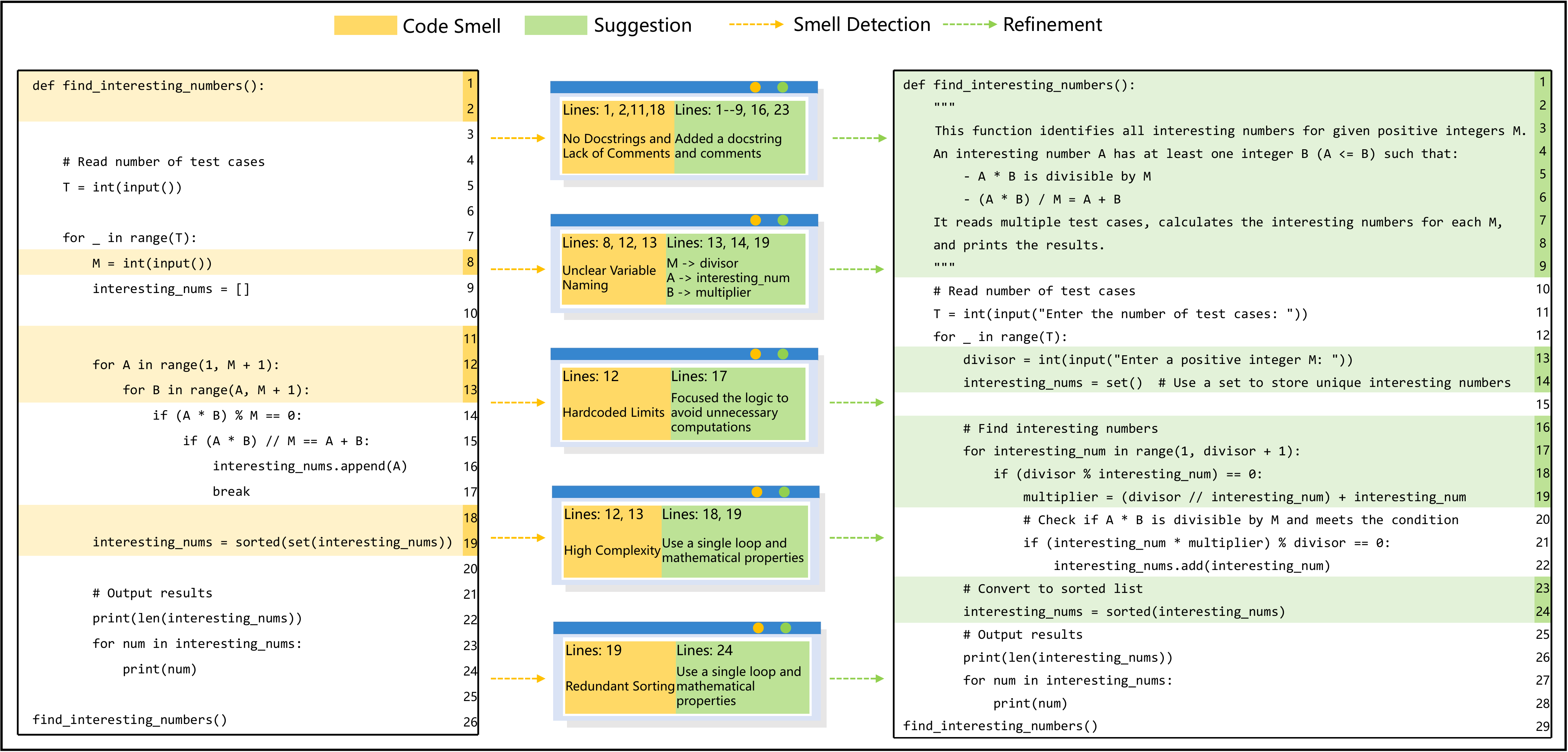}}
\caption{The Code Refinement Mechanism}
\label{fig:code_refinement_mechanism}
\end{figure}

Figure~\ref{fig:code_refinement_mechanism} illustrates the mechanism of self-refinement for a instance in the test dataset. It can be seen that the workflow of this mechanism begins with Code Smell Analysis for smell detection, identifying existing issues such as No Docstrings and Lack of Comments, Unclear Variable Naming, Hardcoded Limits, High Complexity, and Redundant Sorting. Subsequently, the built-in AI system proposes corresponding improvements for these suggestions. Finally, these suggestions are used as system prompts to achieve self-refinement.

\begin{figure}[htbp]
\centerline{\includegraphics[width=0.6\textwidth]{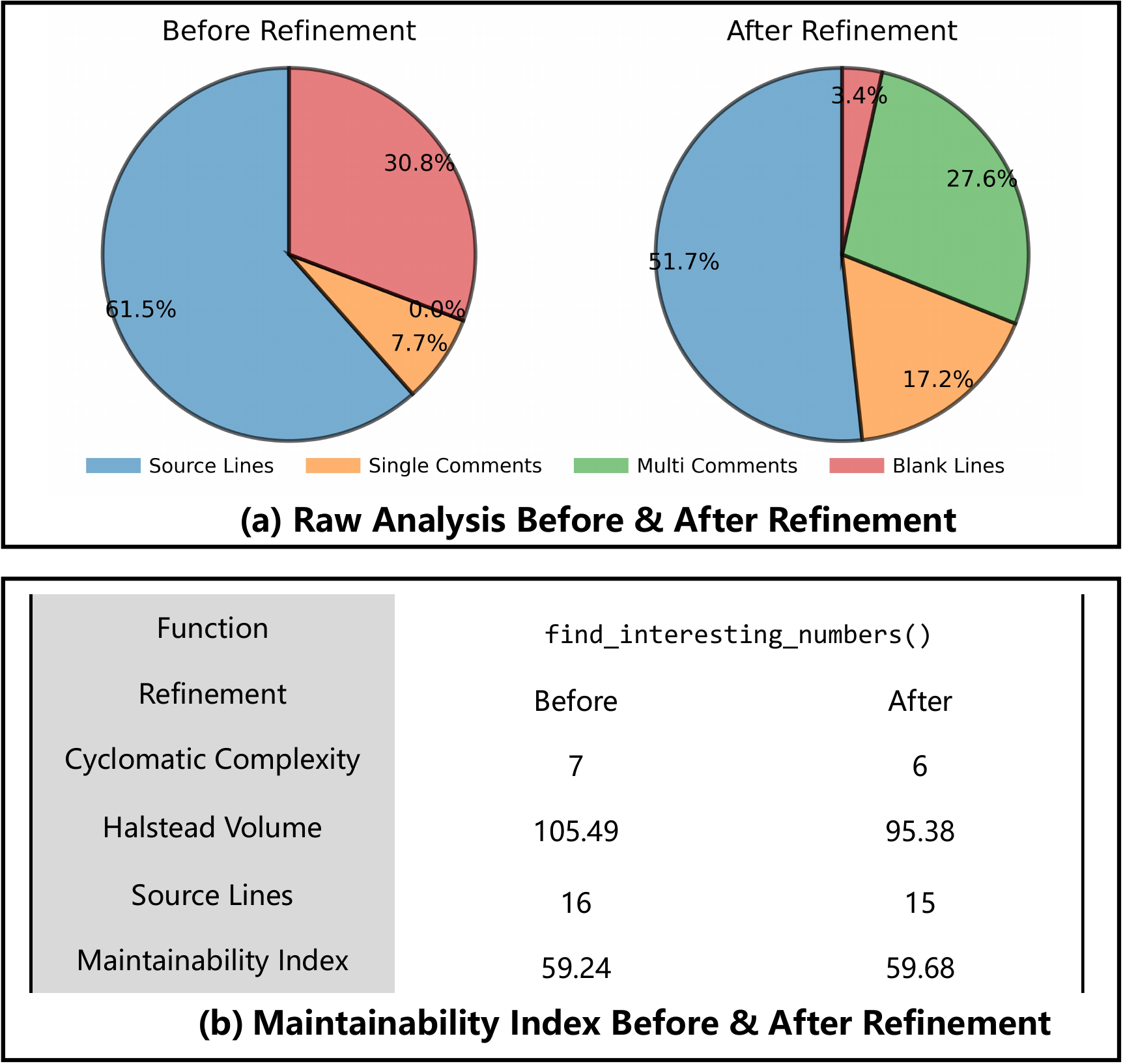}}
\caption{Code Maintainability Report Before and After Refinement
}
\label{fig:matintainability_before_after}
\end{figure}

Figure~\ref{fig:matintainability_before_after} shows the results of analyzing certain metrics of the code before and after refinement using maintainability analysis, quantifying the effectiveness of the refinement. As can be seen from Figure~\ref{fig:matintainability_before_after}(a), there is a significant increase in the ratio of single-line and multi-line comments in the code compared to before, which undoubtedly enhances the interpretability of the code. Furthermore, as shown in Figure~\ref{fig:matintainability_before_after}(b), after refinement, the overall cyclomatic complexity and Halstead Volume of the code have decreased, resulting in an increase in the Maintainability Index compared to before refinement, further demonstrating the positive feedback of the entire refinement on code maintainability.

\section{Additional Application Scenarios}
\label{appendix:casestudy}

To further demonstrate the effectiveness and efficiency of our tool, we present it in three downstream code-related tasks. 
Our tool can significantly streamline LLM workflows in these tasks, greatly reducing developers' time and costs.

\subsection{Unit Test Generation}

\begin{figure}[htbp]
\centerline{\includegraphics[width=1\textwidth]{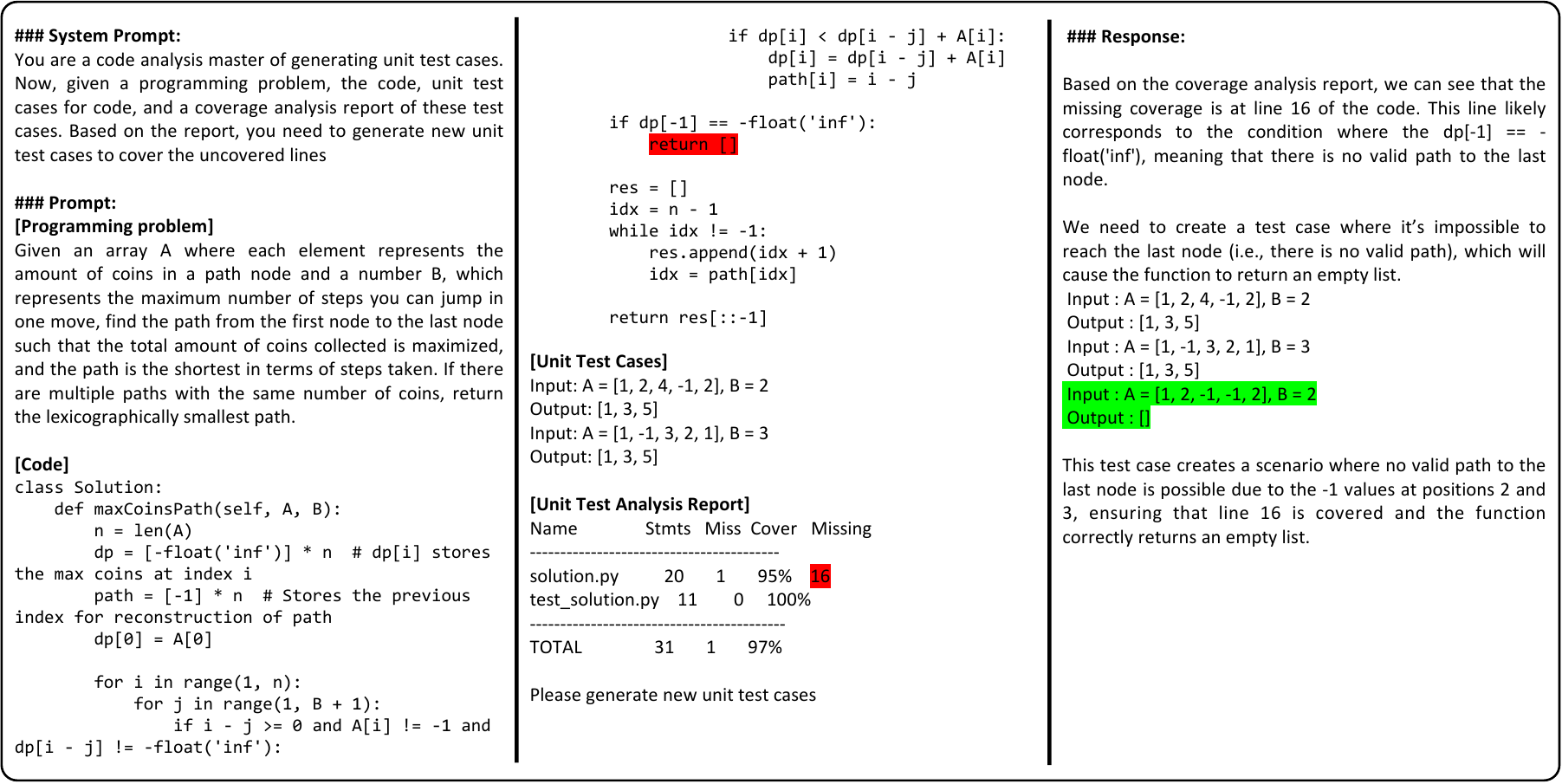}}
\caption{Unit test generation through MPLSandbox}
\label{fig:instance1}
\end{figure}

In this experiment, we show how to generate more effective unit test samples through MPLSandbox to evaluate the user's code.
User-provided unit test samples may not cover the entire code, \ie the unit test coverage is not 100\%. 
This means there could be errors in code snippets that have not been verified. 
Following prior work \cite{jiang2024towards}, we can utilize the unit test analysis tool to identify the uncovered code snippets. 
We then integrate these snippets into the prompt to enable the LLM to generate additional unit test samples to verify these uncovered snippets.
Through MPLSandbox, unit test generation is simple and effective.
Users can modify the system prompt to achieve this task easily.
An instance is illustrated in Figure~\ref{fig:instance1}.

\subsection{Code Translation}

Second, we showcase translating code from one programming language to another.
Large language models have been widely used in code translation. 
Prior work \cite{tao2024unraveling, luo2024bridging} has demonstrated that including structural information of the code, such as unit test cases and the control flow graph, into the system prompt can help LLMs better understand the code, significantly increasing the success rate of code translation.

With MPLSandbox, users can effortlessly integrate various code-related information from traditional analysis tools into the system prompt, to enhance the performance of code translation.
An instance of translating code from Python into Java is shown in Figure~\ref{fig:instance1}.

\begin{figure}[htbp]
\centerline{\includegraphics[width=1\textwidth]{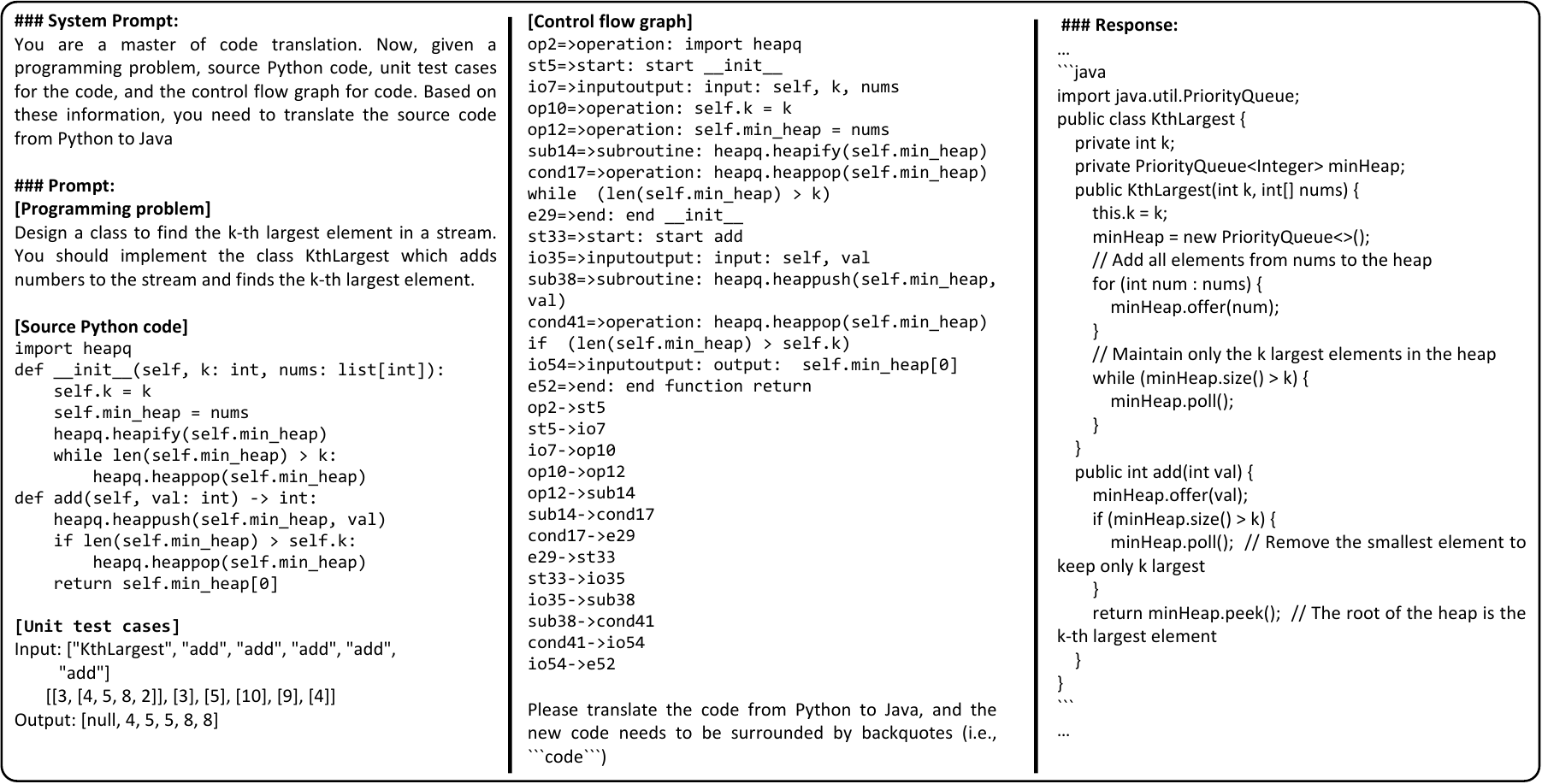}}
\caption{Code translation through MPLSandbox}
\label{fig:instance2}
\end{figure}

\subsection{Vulnerability Location}

\begin{figure}[htbp]
\centerline{\includegraphics[width=1\textwidth]{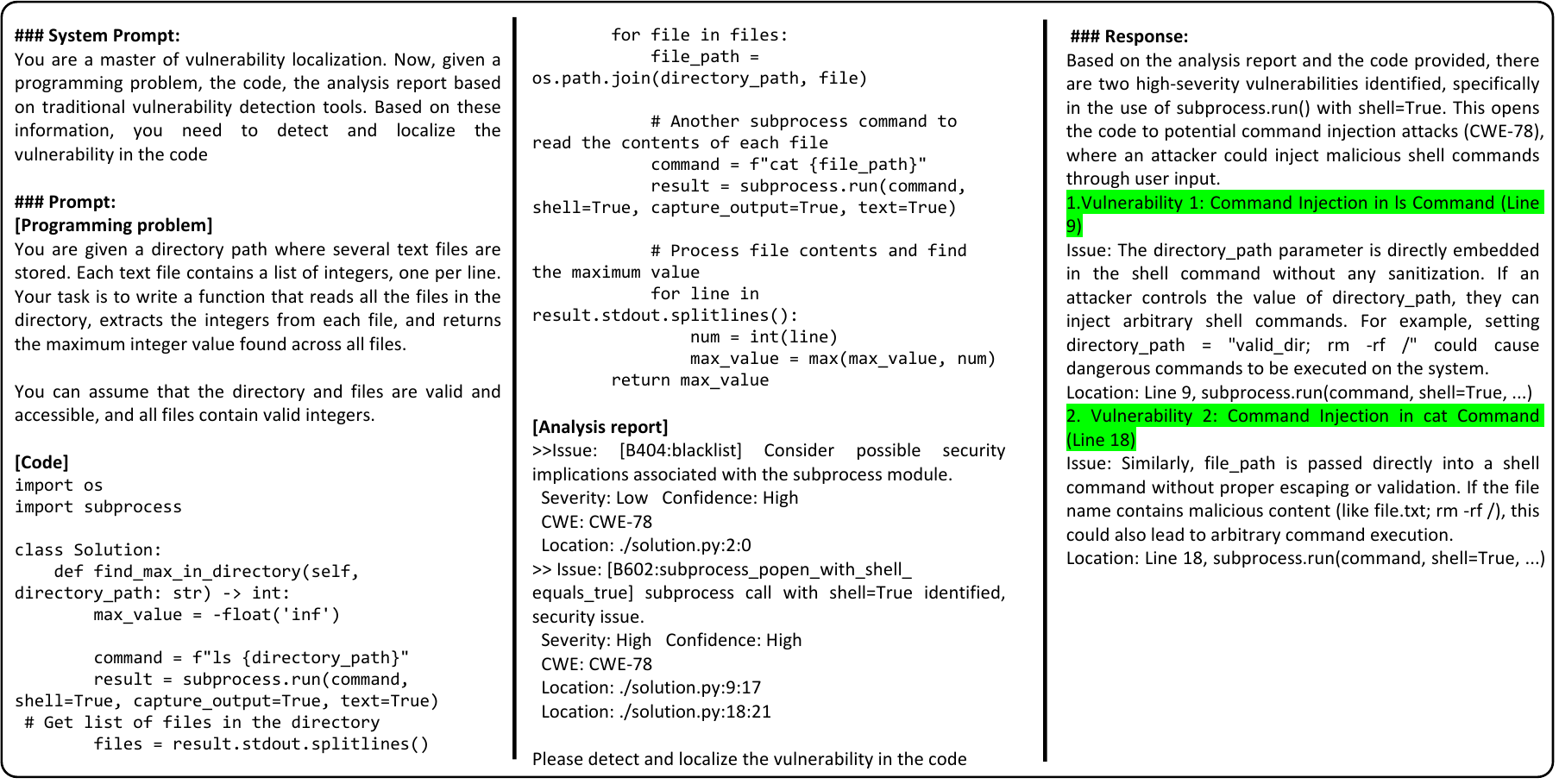}}
\caption{Vulnerability location through MPLSandbox}
\label{fig:instance3}
\end{figure}

LLMs can assist developers in identifying potential vulnerabilities in their code. 
When combined with traditional vulnerability analysis tools, LLMs can significantly enhance the performance of vulnerability detection and localization \cite{lu2024grace, akuthota2023vulnerability}. 
MPLSandbox can also efficiently deploy LLM-based vulnerability detection workflows, minimizing development costs.
An instance of detecting vulnerabilities in the code through MPLSandbox is illustrated in Figure~\ref{fig:instance3}.

In summary, MPLSandbox can enhance the efficiency of deploying LLMs in various code-related tasks.

\end{document}